\documentclass[10pt,journal,compsoc]{IEEEtran}
\ifCLASSOPTIONcompsoc
  \usepackage[nocompress]{cite}
\else
  \usepackage{cite}
\fi
\usepackage[ruled,vlined,lined,commentsnumbered]{algorithm2e}
\usepackage{amsmath}
\usepackage{wasysym}
\usepackage{booktabs}
\usepackage{moreverb}
\usepackage{fontenc}
\usepackage{amsmath}

\usepackage{amssymb}
\usepackage{fancybox}
\usepackage{color}
\usepackage{colortbl}
\usepackage{array}
\usepackage{multirow}
\usepackage{multicol}
\usepackage{listings}
\usepackage{makecell}
\usepackage{graphicx}
\usepackage{setspace}
\usepackage{soul}
\usepackage[breaklinks]{hyperref}
\usepackage{balance}
\usepackage{csvsimple}
\usepackage{longtable}
\usepackage[skip=0pt,font=small,labelfont=bf]{caption}
\usepackage{url}
\usepackage{lscape}
\usepackage{rotating}
\usepackage{tikz}
\usepackage[normalem]{ulem}
\usepackage{enumitem}
\usepackage{subcaption}
\usepackage{wrapfig}
\usepackage[most]{tcolorbox}
\usepackage{marvosym}
\usepackage{pifont}%
\usepackage{threeparttable}
\usepackage{marvosym}
\usepackage{pifont}

\newcolumntype{L}[1]{>{\raggedright\let\newline\\\arraybackslash\hspace{0pt}}m{#1}}
\newcolumntype{C}[1]{>{\centering\let\newline\\\arraybackslash\hspace{0pt}}m{#1}}
\newcolumntype{R}[1]{>{\raggedleft\let\newline\\\arraybackslash\hspace{0pt}}m{#1}}

\definecolor{codeblack}{rgb}{0,0,0}
\definecolor{codeblue}{rgb}{0,0,1}
\definecolor{codegreen}{rgb}{0,0.6,0}
\definecolor{codegray}{rgb}{0.5,0.5,0.5}
\definecolor{codepurple}{rgb}{0.58,0,0.82}
\definecolor{backcolour}{rgb}{0.95,0.95,0.92}
\definecolor{lightgreen}{HTML}{99d8c9}
\definecolor{darkred}{HTML}{760c04}

\lstdefinestyle{mystyle}{
    commentstyle=\color{codegreen},
    keywordstyle=\color{magenta},
    numberstyle=\tiny\color{black},
    stringstyle=\color{codepurple},
    basicstyle=\footnotesize,
    breakatwhitespace=false,
    breaklines=true,
    captionpos=b,
    keepspaces=true,
    showspaces=false,
    showstringspaces=false,
    showtabs=false,
    tabsize=2
}

\setlist{noitemsep} 
\newcommand{\myscriptsize}{\fontsize{6.5pt}{6pt}\selectfont}
\definecolor{darkpastelred}{rgb}{0.76, 0.23, 0.13}
\definecolor{ao(english)}{rgb}{0.0, 0.5, 0.0}

\definecolor{darkpastelred}{rgb}{0.76, 0.23, 0.13}
\definecolor{ao(english)}{rgb}{0.0, 0.5, 0.0}
\lstdefinelanguage{diff}{
	morecomment=[f][\color{blue}]{@@},     
	morecomment=[f][\color{red}]-,         
	morecomment=[f][\color{codegreen}]+,       
	morecomment=[f][\color{red}]{---}, 
	morecomment=[f][\color{codegreen}]{+++},
}

\definecolor{yellow}{RGB}{255,255,153}
\definecolor{grey}{RGB}{224,224,224}

\newboolean{showcomments}
\setboolean{showcomments}{true}
\ifthenelse{\boolean{showcomments}}
 { \newcommand{\mynote}[2]{
      \fbox{\bfseries\sffamily\scriptsize#1}
        {\small$\blacktriangleright$\textsf{\emph{#2}}$\blacktriangleleft$}}}
        { \newcommand{\mynote}[2]{}}

\definecolor{DarkOrange}{rgb}{0.8,0.3,0.0}
\definecolor{DarkCyan}{rgb}{0.0, 0.55, 0.55}

\newcolumntype{?}{!{\vrule width 1pt}}

\definecolor{grey}{rgb}{0.9,0.9,0.9}
\definecolor{lightgrey}{HTML}{f0f0f0}
\definecolor{mygreen}{HTML}{02818a}
\definecolor{mygray}{HTML}{666666}

\newcommand{\toolname}{\texttt{TemVUR}\xspace}
\newcommand{\datasetname}{ManyVuls4J\xspace}
\newcommand*{\eg}{e.g., }

\newcommand*{\ie}{i.e., }

\newcommand*{\etal}{et al. }

\ifCLASSINFOpdf
\else
\fi
\newcommand{\fixpattern}[1]{
\vspace{-0.15cm}
\begin{tcolorbox}[tile,size=fbox,boxsep=0mm,boxrule=0pt,top=0pt,bottom=0pt,
borderline west={2mm}{0pt}{black!5!white},colback=black!5!white] 
#1
\end{tcolorbox}
\vspace{-0.15cm}
}

\lstdefinelanguage{pattern}{
  language=Java,
    backgroundcolor=\color{gray!10!white},   
    commentstyle=\color{green},
    keywordstyle=\color{red},
    numberstyle=\tiny\color{gray},
    stringstyle=\color{red},
    basicstyle=\ttfamily\footnotesize,
    breakatwhitespace=false,         
    breaklines=true,                 
    captionpos=b,                    
    keepspaces=true,                 
    numbersep=5pt,                  
    showspaces=false,                
    showstringspaces=false,
    showtabs=false,                  
    tabsize=2,
    morecomment=[s][\color{green!80!blue}]{+\ },
    morecomment=[s][\color{red}]{-\ },
    morecomment=[n][\color{violet}]{<mask}{>},
    escapeinside={(*}{*)},
}

\begin{document}
%
\title{There are More Fish in the Sea: Automated Vulnerability Repair via Binary Templates}
%
%
%
%

\author{
        Bo~Lin,
        Shangwen~Wang,
        Liqian~Chen
        and~Xiaoguang~Mao 
\IEEEcompsocitemizethanks{\IEEEcompsocthanksitem Bo Lin, Shangwen Wang, Liqian Chen and Xiaoguang Mao are with the Key Laboratory of Software Engineering for Complex Systems, College of Computer Science and Technology, National University of Defense Technology, Changsha, China. \\ E-mails: linbo19@nudt.edu.cn, wangshangwen13@nudt.edu.cn, lqchen@nudt.edu.cn and xgmao@nudt.edu.cn \protect

 
 
 
 \IEEEcompsocthanksitem Shangwen Wang is the corresponding author.
 }
}
\IEEEtitleabstractindextext{%
\begin{abstract} 
As software vulnerabilities increase in both volume and complexity, vendors often struggle to repair them promptly.
Automated vulnerability repair has emerged as a promising solution to reduce the burden of manual debugging and fixing activities. However, existing techniques exclusively focus on repairing the vulnerabilities at the source code level, which has various limitations. For example, they are not applicable to those (\eg users or security analysts) who do not have access to the source code. Consequently, this restricts the practical application of these techniques, especially in cases where vendors are unable to provide timely patches.
In this paper, we aim to address the above limitations by performing vulnerability repair at binary code level, and accordingly propose a template-based automated vulnerability repair approach for Java binaries. 
Built on top of the literature, we collect fix templates from both existing template-based automated program repair approaches and vulnerability-specific analyses, which are then implemented for the Java binaries.
Our systematic application of these templates effectively mitigates vulnerabilities: experiments on the Vul4J dataset demonstrate that \toolname successfully repairs 11 vulnerabilities, marking a notable 57.1\% improvement over current repair techniques. 
Moreover, \toolname securely fixes 66.7\% more vulnerabilities compared to leading techniques (15 vs. 9), underscoring its effectiveness in mitigating the risks posed by these vulnerabilities.
To assess the generalizability of \toolname, we curate the \datasetname dataset, which goes beyond Vul4J to encompass a wider diversity of vulnerabilities. With 30\% more vulnerabilities than its predecessor (increasing from 79 to 103), \datasetname serves as a robust benchmark for evaluating the performance of \toolname. The evaluation on \datasetname reaffirms \toolname's effectiveness and generalizability across a diverse set of real-world vulnerabilities.

\end{abstract}

}

\maketitle

\IEEEdisplaynontitleabstractindextext

%
\IEEEpeerreviewmaketitle


%
%
%
%

\section{Introduction}
\label{sec:intro}
Modern software systems are witnessing an unprecedented surge in complexity and scale, a phenomenon underscored by the escalating prevalence of software vulnerabilities~\cite{radjenovic2013software, homaei2017seven}. In this scenery, detecting and patching software vulnerabilities are crucial activities in the software development industry. 
According to the statistics of the Common Vulnerabilities and Exposures, the number of vulnerabilities discovered per year is considerably increased four-fold from 6,000+ per year in 2016 to 25,000+ per year in 2022~\cite{cve2023}. With the increasing number of vulnerabilities, the time to fix the vulnerability also increases. According to the report from Veracode~\cite{veracode2022security}, 68\% of the vulnerabilities are not fixed within the first three months. In this context, the silver lining is represented by automation, especially by automated tools for vulnerability repair.

Recently, researchers have proposed various approaches to help developers understand the characteristics of vulnerabilities~\cite{bui2024apr4vul,lin2020software,bosu2014identifying} and find vulnerabilities faster~\cite{chakraborty2021deep,fu2022linevul,hin2022linevd,morrison2018vulnerabilities,nguyen2019deep,croft2022data,moustapha2022active}. 
Such techniques help developers analyze, detect, and localize vulnerabilities, but the developers still have to spend lots of manual efforts fixing the vulnerability. Automated Vulnerability Repair (AVR) is regarded as a promising techniques that can alleviate the heavy burden of manual debugging and fixing activities. Specifically, some researchers proposed to repair program vulnerability via program analysis techniques~\cite{zhang2022program, gao2021beyond, huang2019using}. For instance, Gao \etal~\cite{zhang2022program} posed a counter-example guided inductive inference procedure to define likely invariants and use them to construct patches via simple patch templates.
Recently, the adoption of advanced deep learning techniques for vulnerability repair has emerged as a notable trend~\cite{chi2022seqtrans,fu2022vulrepair,chen2022neural}. For instance, VulRepair~\cite{fu2022vulrepair} utilize the pre-trained model, along with the domain knowledge, to repair vulnerabilities.

In the software supply chain, software usually has various forms for different participants. For instance, vendors typically handle software in the form of source code, while downstream users only have compiled binary code, as it is the most common form of software distribution. Despite that a number of program analysis-based~\cite{gao2021beyond,zhang2022program,lee2018memfix} and history-driven AVR techniques~\cite{chen2022neural,fu2022vulrepair,chi2022seqtrans} have been proposed during the years, all of them exclusively target repairing vulnerabilities at the source code level. To our best knowledge, there is no known binary-level AVR technique in the literature. Automating the repair process on the vendor side through source code-level vulnerability fixes, although effective, may encounter significant limitations considering the following two factors:
\begin{itemize}[leftmargin=*]
    \item {\bf Unavailability of source code}. 
    AVR techniques designed for source code are not applicable for those with no access to the source code, such as downstream users or security analysts, as some vendors often encrypt or obfuscate their software, which is particularly prevalent within commercial off-the-shelf software~\cite{duck2020binary}.
    Given that vendors often do not promptly repair the vulnerabilities~\cite{veracode2022security}, the software users may need to actively repair the vulnerabilities they encounter during the execution of software to avoid malicious attacks. In terms of this, existing AVR techniques bring negligible benefits to this process.
    \item {\bf Inefficiency}: Existing vulnerability repair techniques have to compile and load each candidate patch during the patch generation process. Compiling a program can take a significant amount of time. For instance, the compilation and validation of patches can consume up to 92.8\% of the total execution time of automated program repair tools~\cite{chen2017contract}. Considering that vendors would have tens of thousands of vulnerabilities to deal with~\cite{cve2023}, such a process would be rather time-consuming and unaffordable.
\end{itemize}
The aforementioned limitations of fixing vulnerabilities at the source code level emphasize the need for a fresh approach to vulnerability repair.
While decompilation can mitigate the unavailability of source code, it is challenging to obtain the executable code due to software encryption and missing dependencies~\cite{duck2020binary}. Therefore, we foresee the binary code level vulnerability repair as a promising way to alleviate such limitations. On one hand, fixing vulnerabilities at the binary code level would enable users or security analysts to deal with encountered vulnerabilities in a timely manner; on the other hand, it would also enhance the repair efficiency of vendors, since the extremely time-consuming compilation process can be skipped if the modification is performed on the binaries. 


Motivated by this, in this paper, we propose \toolname, an automated vulnerability repair approach for Java binaries (\ie bytecode files), aiming to mitigate the aforementioned limitations of existing source code-level AVR tools. 
\toolname is a \underline{tem}plate-based \underline{vu}lnerability \underline{r}epairor which works by pre-defining several repair templates and then matching the binaries with such templates and taking corresponding repair actions. Such a technical design is inspired by the fact that the amount of vulnerability-related data is considerably small, which implies that data-driven approaches such as language models \cite{fu2022vulrepair,chi2022seqtrans} would not have abundant data for sufficient training. For instance, the training set of Seqtrans~\cite{chi2022seqtrans} only includes 708 vulnerabilities along with their corresponding patches.
To build the template set, we first systematically review the AVR and APR literature, collecting repair templates from related papers.
We then manually assess the templates and categorize them based on their repair actions. 
In total, we collect 33 templates across 14 categories, among which 13 templates are vulnerability-specific (\ie collected from the AVR literature) while the others are reused from the existing APR techniques. For example, the collected templates can mitigate the usage of insecure APIs as well as prevent invalid access such as improper array indexes, both of which are formulated as fixing templates for the first time (details could be referred to Section~\ref{sec:method}).
Given a vulnerable program, our tool begins by performing fault localization, generating a list of suspicious instructions along with their respective suspiciousness scores. We then select the appropriate fix templates for these suspicious instructions and generate candidate patches. Finally, we use the test suites as the oracle to check the generated patches. \toolname returns the patches that pass all test suites for further human review.


We conducted extensive experiments comparing \toolname with contemporary APR and AVR approaches using the widely-adopted Vul4J benchmark~\cite{bui2022vul4j}. We focus on the effectiveness of \toolname in correctly repairing and securely fixing vulnerabilities. A correct repair means the generated patch is semantically equivalent to the developer's patch. A security-fixing patch refers to a patch that prevents the exploitation of vulnerabilities, regardless of whether it corrects the vulnerability.
The results indicate that \toolname is simple yet effective that outperforms all other examined APR and AVR techniques. It increases the number of correctly fixed vulnerabilities from 7 to 11, and securely fixes 66.7\% more vulnerabilities (9 $\rightarrow$ 15) compared to leading techniques. Furthermore, \toolname correctly fixes two and securely fixes three vulnerabilities that no previous tools achieved.
To evaluate the generalizability of \toolname, we constructed a \datasetname using the identical selection criteria and data source (\ie ProjectKB~\cite{ponta2019msr}) as Vul4J, but with the most recent version of ProjectKB (detailed in Section~\ref{subsec:dataset}). Consequently, we expanded the number of vulnerabilities in Vul4J from 79 to 103, which is 30\% larger than the original version. The evaluation performed in \datasetname indicates that \toolname maintains a similar performance in \datasetname, demonstrating the generalizability of \toolname.
To sum up, the contributions of this paper are as follows:
\begin{itemize}[leftmargin=*]
    \item {\bf New Dimension:} Our work focuses on preventing the exploitation of vulnerabilities at the binary level which is neglected by previous AVR techniques. Getting rid of the dependence on source code, our tool has a wider range of application scenarios.
    
    \item {\bf State-of-the-art AVR tool:} We propose a template-based AVR tool called \toolname, which achieve the state-of-the-art performance on vulnerability repair. Moreover, \toolname could also serve as a baseline for future AVR tools through demonstrating the performance a straightforward approach can achieve.
    
    \item {\bf New Dataset:} Datasets are vital to the vulnerability repair community, as they offer a platform for assessing tools' performances. We created a dataset named \datasetname, which is approximately 30\% larger than the existing largest one. Each vulnerability in \datasetname is gathered from real-world instances and has undergone manual verification to ensure its reproducibility.

    \item {\bf Available artifacts:} To support the community, we release materials such as replication package, generated patches, and the dataset used in our experiments for replication and future studies at \url{https://zenodo.org/records/11084782}.
\end{itemize}


\section{Background and Related Works}
\label{sec:bg}

\subsection{Software Vulnerabilities}
A software vulnerability refers to a weakness in the code that can be leveraged by an attacker to carry out unauthorized actions. For instance, the `Out-of-bounds Write', ranked as the top stubborn weakness in the Common Weakness Enumeration database, enables the attacker to inject or execute malicious code. Vulnerabilities are exploited over a million times each year, costing the global economy hundreds of billions of dollars due to cybercrime~\cite{armin20152020}.

Common Vulnerabilities and Exposures (CVE) is a popular database that collect thousand of expose vulnerability every month. Each vulnerability would be assigned a unique CVE ID as the identifier. For example, Dirty COW is a privilege escalation vulnerability in the Linux kernel, identified as CVE-2016-5195. The year 2016 indicates the time this vulnerability was exposed, and 5,195 uniquely identifies the vulnerability within that year.

Common Weakness Enumeration (CWE) is a category for vulnerabilities. Each vulnerability in CVE with a unique CVE ID would be categorized into a CWE category representing the generic type of this vulnerability. For example, CWE-787 represents the `Out-of-bounds Write' category, while the Dirty COW is assigned to CWE-362, the `Race Condition' category. As of February 2024, the total number of vulnerability categories in CWE has reached 938~\cite{cwe2024}.

\subsection{Automated Vulnerability Repair}
Since the inception of Genprog~\cite{le2012genprog}, numerous of Automated Program Repair (APR) techniques have emerged~\cite{yuan2020arja,liu2019tbar,lutellier2020coconut,zhang2023gamma}. These techniques significantly reduce manual debugging efforts, enabling automated patch generation. 
However, APR techniques primarily target general bugs, showing a discrepancy with vulnerabilities. 
General bugs refers to an unintended error or flow in the code that disrupt the intended functionality of a program, potentially causing the program to behave unpredictably or crash. In contrast, a vulnerability denotes a weakness in the design, implementation or configuration of a software that can be exploited by malicious actors to compromise its security.
This gap between general bugs and vulnerabilities leads to poor performance of APR techniques in vulnerability repair~\cite{bui2024apr4vul,pinconschi2021comparative}. AVR techniques, on the other hand, prioritize fixing vulnerabilities over general bugs. According to the study from Pinconschi \etal ~\cite{pinconschi2021comparative}, AVR techniques can be categorized based on their design target: those that aim to repair specific types of vulnerabilities, and those designed for general vulnerabilities.

Target-specific AVR techniques focus on specific security vulnerabilities, such as buffer overflow or null pointer. These vulnerabilities share common features: they have a clear root cause, and the process of localization and patch generation can be formalized. For instance, the typical root cause of a buffer overflow vulnerability is a program trying to read or write data beyond the buffer boundary. Most of existing studies utilized template~\cite{sidiroglou2015automatic,sidiroglou2015automatic2,gao2020automatic} and program analysis~\cite{huang2019using,gao2021beyond} to repair this type of vulnerability. Gao \etal, for instance, utilize a sanitizer to extract a constraint, which severs as the proof obligation that the synthesized patch should satisfy. After that, the KLEE symbolic execution engine~\cite{cadar2008klee} is utilized to generate the patches. Target-specific AVR techniques have achieved inspiring results. However, the design and implementation of these techniques require a significant amount of manual efforts. As for February 2024, CWE~\cite{cwe2024} reports 938 vulnerability categories, indicating that it is challenging for researchers to develop tools to cover all the kinds of vulnerabilities.

To mitigate the limitations of target-specific AVR techniques, some researcher proposed AVR techniques that are not restricted by specific vulnerability types~\cite{ma2017vurle,harer2018learning,chi2022seqtrans, chen2022neural, fu2022vulrepair}. Most of these techniques take the advantage of deep learning techniques to learn the repair action from the history vulnerability fixing commits. Typical general target AVR techniques formulate the repair of vulnerabilities as a neural machine translation task. This aims to learn the mapping between a vulnerable program and a fixed version of that program. Such AVR techniques do not need to consider the repair strategy for each type of vulnerability and can perform an end-to-end repair. Recent years, with the aid of deep learning techniques, AVR techniques in this domain have achieved promising results.
However, since all repair actions are learned from historical data, the quantity and diversity of historical data directly impacts the effectiveness of such AVR techniques. The largest existing dataset contains only 5,495 vulnerability fixing commits~\cite{bhandari2021cvefixes}, which could potentially limit the performance of these techniques. While existing AVR techniques have achieved promising results, they primarily focus on repairing vulnerabilities at the source code level.
This limits the application scenarios for AVR tools and underscores the importance of repairing the vulnerabilities at the binary level, as the source code is not always available for users.

\subsection{Java Binary}
Java binary, also known as bytecode, serves as an intermediary between Java source code and machine code executed on the Java Virtual Machine (JVM), bridging the gap between source code and machine instructions. This intermediate form allows Java programs to be platform-independent since they are not compiled into native machine code, but into a format universally executable across different JVM implementations, resulting in the creation of a {\tt class} file for each compiled Java class. These files contains various information such as the superclass, field declarations, and method declarations. Method implementations are stored as a sequence of bytecode instructions. Each bytecode instruction consists of one opcode and zero or more operands. For instance, the load instruction (\eg {\tt aload}) loads a reference onto the stack from a local variable specified by the operand, while the {\tt pop} instruction discards the top value on the stack without needing an operand. The architecture of the bytecode instruction set is a blend of stack-based and register-based machine principles. 
Such as the result value resides on the stack when the invoked method returns. Local variables are represented by a virtually unlimited number of virtual registers stored in the Local Variable Table.
It is noteworthy that although bytecode was initially designed for Java, it now accommodates various programming languages such as Kotlin, Groovy, and Scala, all of which can be compiled to bytecode. This suggests that binary-level vulnerability repair can be broadly applied.
\section{\toolname}
\label{sec:method}
\begin{figure}[!t]
  \centering
  \resizebox{0.98\linewidth}{!}{
    \includegraphics[]{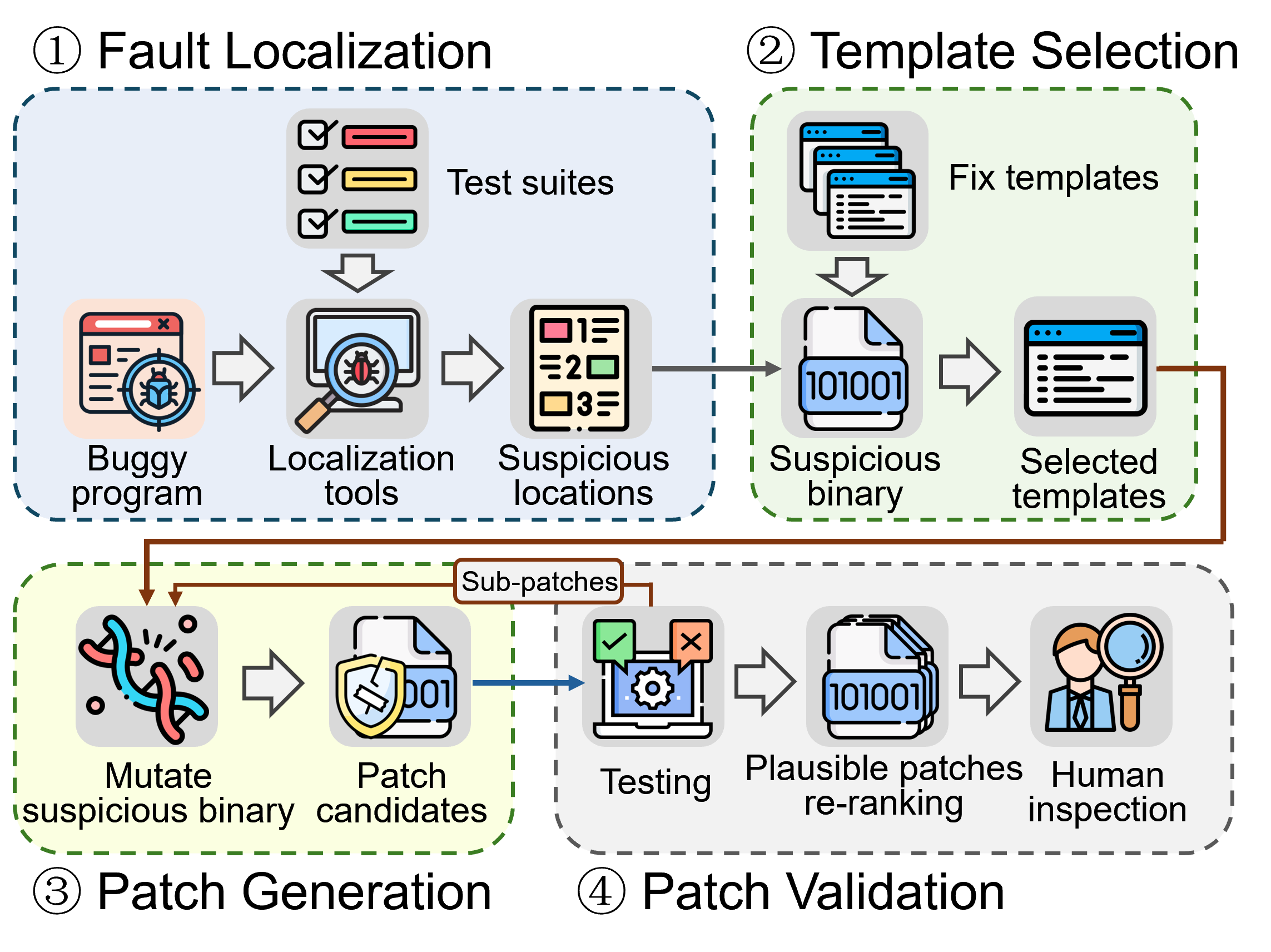}
    }
  \caption{The overall workflow of \toolname.}
  \label{fig:pipeline}%
\end{figure}

To explore the applicability of patterns in Java binary repair, we developed \toolname, a template-based vulnerability repair approach for Java binaries that implements APR and AVR templates at the Java binary level. 
We expect the community to consider \toolname as a baseline to observe the performance achievable through the straightforward application of repair templates. Newly-proposed approaches should introduce innovative techniques to resolve auxiliary issues, such as repair precision and search space optimization. \toolname can serve as a baseline to reflect the improvements brought by these approaches.
Figure~\ref{fig:pipeline} presents the workflow of \toolname. 
Given a program with vulnerabilities and a set of test suites, including at least one failing test case (\ie proof of vulnerability), the FL algorithm supplies a list of suspicious code locations (\eg code lines) along with their respective suspiciousness scores.
We then select the appropriate fix patterns for all suspicious code locations and generate candidate patches. For all generated patches, \toolname employs the test suites as the oracle to check the generated patches and returns plausible patches. 
Finally, we rank the plausible patches for manual inspection and patches that modify less suspicious locations receive a higher rank, in line with previous APR works~\cite{chen2017contract,wen2018context,le2017s3,wang2020automated}.  

\subsection{Fix Template Definition}
\label{subsec:templates}
In the existing literature, various fix templates have been designed based on manual summarization, pre-definition, and automatic mining.
According to a recent study ~\cite{bui2024apr4vul}, most repair actions for general bugs can also be found in vulnerabilities.
Therefore, we collect fixing templates on top of the state-of-the-art template-based APR techniques~\cite{ghanbari2019practical,liu2019tbar}, as well as the empirical study for vulnerabilities~\cite{bui2024apr4vul}, and then implement them in Java binaries. In total, we identified 14 categories of templates labeled according to the repair actions in the pattern. Within each category, there may contain several related patterns.  
In the following, we list details of these templates, where a template ID in \textcolor{darkred}{dark red} (\eg T1.1) indicates that the template is sourced from the AVR literature. Note that patterns implemented in the Java binaries would require more instructions to achieve the target compared with the implementation in the source code. For example, changing a variable's data type from {\tt float} to {\tt double} would involve altering several instruments and adjusting the operand stack. As a result, we provide a binary code example for the first template in each category and use source code to explain the left ones.

{\bf T1. Mutate Object Instantiation.} Mutating the instantiation of an object with null or the secure version. The secure version of a class is either derived from Java documents~\cite{java2017doc} or inferred from accessible classes. We assert that class {\tt B} is the secure version of class {\tt A} if {\tt A} is the parent class of {\tt B} and {\tt B} is named {\tt SecureA}. An example of this is {\tt java.security.SecureRandom}, which is the secure version of {\tt java.util.Random} class.
\fixpattern{
\begin{lstlisting}[language=java,linewidth={\linewidth},basicstyle=\myscriptsize\ttfamily]
(@\phantom{x}{\bf \textcolor{darkred}{T1.1}:}\hspace{2ex} \textcolor{red}{-}\hspace{2ex}@)... = new C();  (@\phantom{x}\hspace{6ex}{\bf T1.1 Bytecode Example:}@)
(@\phantom{x}\hspace{6.7ex} \textcolor{codegreen}{+}\hspace{2ex}@)... = new SecureC();(@\hspace{2.5ex}\textcolor{red}{-}@) new #idx_C
(@\phantom{x}{\bf \textcolor{darkred}{T1.2}:}\hspace{2ex} \textcolor{red}{-}\hspace{2ex}@)... = new C(); (@\phantom{x}\hspace{6.2ex} \textcolor{codegreen}{+}@) new #idx_SecureC
(@\phantom{x}\hspace{6.7ex} \textcolor{codegreen}{+}\hspace{2ex}@)... = null; (@\phantom{x}\hspace{12.4ex} @) dup
(@\phantom{x}\hspace{40.7ex} \textcolor{red}{-}@) invokespecial #idx_C_init
(@\phantom{x}\hspace{40.7ex} \textcolor{codegreen}{+}@) invokespecial #idx_SecureC_init
\end{lstlisting}}
We provide a bytecode example that updates the instantiated of an object with its secure version. Assume that neither of them require constructor parameters. Here, {\tt \#idx\_C} and {\tt \#idx\_SecureC} represent the indexes of {\tt C} and {\tt SecureC} stored in the constant pool. Likewise, {\tt \#idx\_C\_init} and {\tt \#idx\_SecureC\_init} refer to the indexes of the constructor for {\tt C} and {\tt SecureC}, respectively.

{\bf T2. Insert Validity Checker.} Insert a validity checker to a statement that contains an expression which might be invalid, such as a null pointer checker.

\fixpattern{
\begin{lstlisting}[language=java,linewidth={\linewidth},basicstyle=\myscriptsize\ttfamily]
(@\phantom{}\hspace{0ex}{\bf \textcolor{darkred}{T2.1}:}\hspace{2ex}\textcolor{codegreen}{+}@) if ((@\textcolor{codeblue}{check}@)(exp)) {  (@\phantom{}\hspace{12ex}{\bf T2.1 Bytecode Example:}@)
(@\phantom{}\hspace{15ex}@)...exp...; ...... (@\phantom{}\hspace{8.1ex}\textcolor{codegreen}{+}@) aload idx_obj
(@\phantom{}\hspace{5.7ex} \textcolor{codegreen}{+}@) } (@\phantom{}\hspace{34.30ex} \textcolor{codegreen}{+}@) invokevirtual #idx_check
(@\phantom{}{\bf \textcolor{darkred}{T2.2}:}\hspace{2ex}\textcolor{codegreen}{+}@) if ((@\textcolor{codeblue}{check}@)(exp)) (@\phantom{}\hspace{14.6ex} \textcolor{codegreen}{+}@) ifnull LABEL
(@\phantom{}\hspace{13ex}@) return DEFAULT_VALUE;; (@\phantom{}\hspace{1.6ex} \textcolor{codegreen}{ }@) ....
(@\phantom{}\hspace{8.5ex}@) ...exp...; (@\phantom{}\hspace{21.65ex} \textcolor{codegreen}{+}@) LABEL:
(@\phantom{}{\bf \textcolor{darkred}{T2.3}:}\hspace{2ex}\textcolor{codegreen}{+}@) if ((@\textcolor{codeblue}{check}@)(exp)) exp = exp1;  (@\phantom{}\hspace{9.85ex} \textcolor{codegreen}{ }@)
(@\phantom{}\hspace{8.5ex}@) ...exp...;
(@\phantom{}{\bf \textcolor{darkred}{T2.4}:}\hspace{2ex}\textcolor{codegreen}{+}@) if ((@\textcolor{codeblue}{check}@)(exp)) continue;  
(@\phantom{}\hspace{8.5ex}@) ...exp...;
(@\phantom{}{\bf \textcolor{darkred}{T2.5}:}\hspace{2ex}\textcolor{codegreen}{+}@) if ((@\textcolor{codeblue}{check}@)(exp))
(@\phantom{}\hspace{5.7ex} \textcolor{codegreen}{+}\hspace{6ex}@)throw new 
(@\phantom{}\hspace{5.7ex} \textcolor{codegreen}{+}\hspace{6ex}@)IllegalArgumentException(...);  
(@\phantom{}\hspace{8.5ex}@) ...exp...;
\end{lstlisting}}
\noindent
In this context, {\tt check} is a function that verifies whether a given expression complies with predefined rules, and {\tt exp} is an expression or variable. Specifically, for primitive data types (\eg int and float), {\tt check} will evaluate whether {\tt exp} is equal or greater than zero. For arrays and dictionaries, {\tt check} will assess whether {\tt exp} is empty.
For any Java object, {\tt check} will determine whether {\tt exp} is null.
Additionally, {\tt DEFAULT\_VALUE} is the default value as determined by the type of {\tt exp} and the JVM specification~\cite{lindholm2013java}.
The bytecode example demonstrates a fixed template that prevents a null pointer exception by inserting a null pointer check. The {\tt idx\_obj} is the index of the object in the Local
Variable Table (LVT), and the {\tt \#idx\_check} is the index of {\tt check} in the constant pool. 

{\bf T3. Insert Statements.} Templates in this category insert missing statements around the buggy code, including methods invocations, a return statement or try-catch statements.
\fixpattern{
\begin{lstlisting}[language=java,linewidth={\linewidth},basicstyle=\scriptsize\ttfamily]
(@\phantom{x}{\bf T3.1:}\hspace{2ex} \textcolor{codegreen}{+}@)  method(parms);  (@\phantom{x}\hspace{12.1ex}{\bf T3.1 Bytecode Example:}@)
(@\phantom{x}{\bf T3.2:}\hspace{2ex} \textcolor{codegreen}{+}@)  return DEFAULT_VALUE; (@\phantom{x}\hspace{2.4ex} \textcolor{codegreen}{+}@) aload idx_parm
(@\phantom{x}{\bf T3.3:}\hspace{2ex} \textcolor{codegreen}{+}@)  try {                 (@\phantom{x}\hspace{2.4ex} \textcolor{codegreen}{+}@) invokevirtual #idx
(@\phantom{x}\hspace{15ex}@) statement; ......
(@\phantom{x}\hspace{7.1ex} \textcolor{codegreen}{+}@)  } catch (Exception e) { ... }
\end{lstlisting}}
The bytecode example demonstrates the insertion of a method invocation at the bytecode level. This is under the assumption that the {\tt method} in T3.1 is a virtual method and only accepts one parameter.
Here, {\tt idx\_parm} refers to the index of the parameter in the LVT, and {\tt \#idx} corresponds to the index of the {\tt method} in the constant pool.

{\bf T4. Insert Cast Checker.} Adding an instanceof check for a statement that contains an uncheck cast expression.
 \fixpattern{
\begin{lstlisting}[language=java,linewidth={\linewidth},basicstyle=\scriptsize\ttfamily]
(@\phantom{x}\hspace{0ex}{\bf T4.1:}\hspace{2ex} \textcolor{codegreen}{+}@) if (exp instanceof T) {   (@\phantom{x}\hspace{2.1ex}{\bf T4.1 Bytecode Example:}@)
(@\phantom{x}\hspace{14ex}@) var = (T) exp; ...... (@\phantom{x}\hspace{2.1ex} \textcolor{codegreen}{+}@)  aload x
(@\phantom{x}\hspace{7.05ex} \textcolor{codegreen}{+}@)  }  (@\phantom{x}\hspace{31.6ex} \textcolor{codegreen}{+ } \textcolor{codeblack}{instanceof \#idx}@) 
(@\phantom{x}\hspace{49.9ex} \textcolor{codegreen}{+}@)  ifeq LABEL 
(@\phantom{x}\hspace{54.0ex}@) aload (@{\em x}@)
(@\phantom{x}\hspace{54.0ex}@) checkcast #idx
(@\phantom{x}\hspace{54.0ex}@) astore y
(@\phantom{x}\hspace{54.0ex}@) ......
(@\phantom{x}\hspace{49.9ex} \textcolor{codegreen}{+}@)  LABEL:
\end{lstlisting}}
\noindent
where {\tt exp} is an expression (\eg field access and method call) and {\tt T} is the casting type, while ``$\ldots\ldots$'' means the subsequent statements dependent on the variable {\tt var}. In bytecode level template, the {\tt \#idx} is the index of type {\tt T} stored in java constant pool; The {\tt aload x} and {\tt astore y} are the instructions to load {\tt exp} to the top of stack and store the cast {\tt exp} to the {\tt var} respectively. Here, {\tt x} and {\tt y} represent the index position of {\tt exp} and {\tt var} in the local variable array, respectively.
The {\tt LABEL} is the offset of instructions within the method.
Note that casting for primitive types (\eg int and float) does not involve the {\tt checkcast} instruction, and this casting is resolved at compile time. Therefore, we do not consider inserting a cast checker for primitive types.

{\bf T5. Mutate Conditional Block.} Mutating a conditional expression updating, removing, adding a sub-expression and changing the keyword of conditional statement.
\fixpattern{
\begin{lstlisting}[language=java,linewidth={\linewidth},basicstyle=\scriptsize\ttfamily]
(@\phantom{x}{\bf T5.1:}\hspace{2ex} \textcolor{red}{-}@)  ...exp1... (@\phantom{x}\hspace{17.5ex}{\bf T5.1 Bytecode Example:}@)
(@\phantom{x}\hspace{7.0ex} \textcolor{codegreen}{+}@)  ...exp2... (@\phantom{x}\hspace{16.4ex} \textcolor{red}{-}@) aload idx_a
(@\phantom{x}{\bf T5.2:}\hspace{2ex} \textcolor{red}{-}@)  ...exp1 Op exp2... (@\phantom{x}\hspace{5.0ex} \textcolor{codegreen}{+}@) aload idx_b
(@\phantom{x}\hspace{7.0ex} \textcolor{codegreen}{+}@)  ...exp1...     (@\phantom{x}\hspace{13.6ex}@) ifnull LABEL
(@\phantom{x}{\bf T5.3:}\hspace{2ex} \textcolor{red}{-}@)  ...exp1...
(@\phantom{x}\hspace{7.0ex} \textcolor{codegreen}{+}@)  ...exp1 Op exp2...
(@\phantom{x}{\bf \textcolor{darkred}{T5.4}:}\hspace{2ex} \textcolor{red}{-} \hspace{0.05ex} \textcolor{codeblue}{kw1}@)(exp)
(@\phantom{x}\hspace{7.0ex} \textcolor{codegreen}{+} \hspace{0.05ex} \textcolor{codeblue}{kw2}@)(exp)
\end{lstlisting}}
\noindent
where {\tt Op} represents the operator, including arithmetic, assignment, comparison, logical and bitwise operator. {\tt kw1} and {\tt kw2} stand for the keywords as described in the Java Specification~\cite{java2017doc}. 
The bytecode example demonstrates the mutation of {\tt if (a != null)} to {\tt if (b != null)}. Here, {\tt idx\_a} and {\tt idx\_b} represent the indexes of {\tt a} and {\tt b} in the LVT, respectively.

{\bf T6. Mutate Data Type.} Replacing the data type in variable declaration or a cast expression.
\fixpattern{
\begin{lstlisting}[language=java,linewidth={\linewidth},basicstyle=\scriptsize\ttfamily]
(@\phantom{x}{\bf T6.1:}\hspace{2ex} \textcolor{red}{-}@)  ...T1 var ...;  (@\phantom{x}\hspace{12.0ex}{\bf T 6.1 Bytecode Example:}@)
(@\phantom{x}\hspace{7.0ex} \textcolor{codegreen}{+}@)  ...T2 var ...; (@\phantom{x}\hspace{12.1ex} \textcolor{red}{-}@) bipush val
(@\phantom{x}{\bf T6.2:}\hspace{2ex} \textcolor{red}{-}@)  ...(T1) exp...; (@\phantom{x}\hspace{10.6ex} \textcolor{red}{-}@) istore idx_var
(@\phantom{x}\hspace{7.0ex} \textcolor{codegreen}{+}@)  ...(T2) exp...; (@\phantom{x}\hspace{10.7ex} \textcolor{codegreen}{+}@) ldc val
(@\phantom{x}{\bf \textcolor{darkred}{T6.3}:}\hspace{2ex} \textcolor{red}{-}@)  ... exp...;     (@\phantom{x}\hspace{10.7ex} \textcolor{codegreen}{+}@) store idx_var 
(@\phantom{x}\hspace{7.0ex} \textcolor{codegreen}{+}@)  ...(T1) exp...;
\end{lstlisting}}
\noindent
where {\tt T1} and {\tt T2} are two different types in java. {\tt exp} is the expression to be cast. 
Assume that {\tt T1} is {\tt int} and {\tt T2} is {\tt float} in T6.1. The bytecode example demonstrates a mutation that changes the type of {\tt var}. Here, {\tt idx\_var} is the index of {\tt var} in LVT, and {\tt val} is the value to be assigned to {\tt var}. 

{\bf T7. Mutate Literal Expression.} Mutate literals to relevant literals or expressions of the same type, including boolean, string, and number literals. The mutation strategies include inversion (for boolean and number literals), self-increment/decrement (for number literals), and selection from context.
\fixpattern{
\begin{lstlisting}[language=java,linewidth={\linewidth},basicstyle=\scriptsize\ttfamily]
(@\phantom{x}{\bf T7.1:}\hspace{2ex} \textcolor{red}{-}@)  ...literal1...  (@\phantom{x}\hspace{12.0ex}{\bf T7.1 Bytecode Example:}@)
(@\phantom{x}\hspace{7.0ex} \textcolor{codegreen}{+}@)  ...literal2... (@\phantom{x}\hspace{12.0ex} \textcolor{red}{-}@) iconst_0
(@\phantom{x}{\bf T7.2:}\hspace{2ex} \textcolor{red}{-}@)  ...literal1... (@\phantom{x}\hspace{12.0ex} \textcolor{codegreen}{+}@) iconst_1
(@\phantom{x}\hspace{7.0ex} \textcolor{codegreen}{+}@)  ...exp...
\end{lstlisting}}
\noindent
where {\tt literal1} and {\tt literal2} are literals of the same type. The {\tt exp} notation indicates that the expression has the same type as {\tt literal1}.
Assume in T7.1 that {\tt literal1} is {\tt false} and {\tt literal2} is {\tt true}. The bytecode example illustrates that a mutation can change a boolean literal from {\tt false} to {\tt true}.

{\bf T8. Mutate Variables.} Mutate variables to compitable variable or expressions.
\fixpattern{
\begin{lstlisting}[language=java,linewidth={\linewidth},basicstyle=\scriptsize\ttfamily]
(@\phantom{x}{\bf T8.1:}\hspace{2ex} \textcolor{red}{-}@)  ...var1...  (@\phantom{x}\hspace{16.5ex}{\bf T8.1 Bytecode Example:}@)
(@\phantom{x}\hspace{7.0ex} \textcolor{codegreen}{+}@)  ...var2... (@\phantom{x}\hspace{16.5ex} \textcolor{red}{-}@) iload idx_var1
(@\phantom{x}{\bf T8.2:}\hspace{2ex} \textcolor{red}{-}@)  ...var1... (@\phantom{x}\hspace{16.5ex} \textcolor{codegreen}{+}@) iload idx_var2
(@\phantom{x}\hspace{7.0ex} \textcolor{codegreen}{+}@)  ...exp...
\end{lstlisting}}
\noindent
where {\tt var1} is at variable in buggy code element. {\tt var2} and {exp} are a compatible variable and expression of {\tt var1} respectively.
Assume that {\tt var1} and {\tt var2} in T8.1 are both integer variables. The bytecode example illustrates a mutation that replaces {\tt var1} with {\tt var2}. Note that the instruction to load a variable could vary depending on type of operand (\eg {\tt fload} and {\tt aload}).

{\bf T9. Mutate Method Invocation Expression.} This involves modifying the method invocation expression with another one that has a compatible return type and the compatible parameter types. Alternatively, one can replace the argument with a compatible expression or variable.
\fixpattern{
\begin{lstlisting}[language=java,linewidth={\linewidth},basicstyle=\myscriptsize\ttfamily]
(@\phantom{x}{\bf T9.1:}\hspace{2ex} \textcolor{red}{-}@)  ...method1(args)...; (@\phantom{x}\hspace{3.5ex}{\bf T9.1 Bytecode Example:}@) 
(@\phantom{x}\hspace{7.0ex} \textcolor{codegreen}{+}@)  ...method2(args)...; (@\phantom{x}\hspace{3.8ex}\textcolor{red}{-}@) invokevirtual #idx_method1
(@\phantom{x}{\bf T9.2:}\hspace{2ex} \textcolor{red}{-}@)  ...method(arg1)...; (@\phantom{x}\hspace{3.6ex} \textcolor{codegreen}{+}@) invokevirtual #idx_method2
(@\phantom{x}\hspace{7.0ex} \textcolor{codegreen}{+}@)  ...method(arg2)...;
\end{lstlisting}}
\noindent
where {\tt method1} and {\tt method2} are method invocation expression. {\tt args}, {\tt arg1}, {\tt arg2} represent the argument(s) in the method invocation. 
Assume that {\tt method1} and {\tt method2} are both virtual methods. In the bytecode example, their indexes in the constant pool are represented as {\tt \#idx\_method1} and {\tt \#idx\_method2} respectively.

{\bf T10. Mutate Operators.}
\fixpattern{
\begin{lstlisting}[language=java,linewidth={\linewidth},basicstyle=\scriptsize\ttfamily]
(@\phantom{x}{\bf T10.1:}\hspace{2ex} \textcolor{red}{-}@)  ...exp1 Op1 exp2... (@\phantom{x}\hspace{6.5ex}{\bf T10.1 Bytecode Example:}@) 
(@\phantom{x}\hspace{8.2ex} \textcolor{codegreen}{+}@)  ...exp1 Op2 exp2... (@\phantom{x}\hspace{5.4ex} \textcolor{red}{-}@) if_icmpne LABEL
(@\phantom{x}{\bf T10.2:}\hspace{2ex} \textcolor{red}{-}@)  ...var Op1...      (@\phantom{x}\hspace{6.5ex} \textcolor{codegreen}{+}@) if_icmpeq LABEL
(@\phantom{x}\hspace{8.2ex} \textcolor{codegreen}{+}@)  ...var Op2...
(@\phantom{x}{\bf \textcolor{darkred}{T10.3}:}\hspace{2ex} \textcolor{red}{-}@)  ...Op1 var...
(@\phantom{x}\hspace{8.2ex} \textcolor{codegreen}{+}@)  ...Op2 var...
\end{lstlisting}}
\noindent
where {\tt var1} and {\tt var2} denote expressions and {\tt Op} is the associated operator. {\tt var} represents the variable (\eg integer and boolean variable) in the buggy code element. The bytecode example assumes that both {\tt exp1} and {\tt exp2} are integers, {\tt Op1} is ``{\tt ==}'' and {\tt Op2} is ``{\tt != }''.

{\bf T11. Remove Statements.} Directly deleting one or more buggy statements.
\fixpattern{
\begin{lstlisting}[language=java,linewidth={\linewidth},basicstyle=\scriptsize\ttfamily]
(@\phantom{x}{\bf T11.1:}\hspace{2ex} \textcolor{red}{-}@)  ...statement;... (@\phantom{x}\hspace{9.9ex}{\bf T11.1 Bytecode Example:}@) 
(@\phantom{x}\hspace{47.7ex} \textcolor{red}{-}@) ifInst LABEL
(@\phantom{x}\hspace{47.7ex} \textcolor{codegreen}{+}@) goto LABEL
\end{lstlisting}}
\noindent
where {\tt statement} represents the removed buggy statement(s). 
The bytecode example demonstrates the removal of an {\tt if} statement, where {\tt ifInst} represents one of the {\tt if} instructions in the JVM specification, such as {\tt if\_acmpeq}.

{\bf T12. Mutate Return Statements.} Replacing the expression in return statements with compatible type expression.
\fixpattern{
\begin{lstlisting}[language=java,linewidth={\linewidth}, basicstyle=\scriptsize\ttfamily]
(@\phantom{x}{\bf T12.1:}\hspace{2ex} \textcolor{red}{-}@)   return exp1; (@\phantom{x}\hspace{13.1ex}{\bf T12.1 Bytecode Example:}@)
(@\phantom{x}\hspace{8.3ex} \textcolor{codegreen}{+}@)   return exp2; (@\phantom{x}\hspace{12.2ex} \textcolor{red}{-}@) aload x
(@\phantom{x}\hspace{47.2ex} \textcolor{codegreen}{+}@) aload y 
(@\phantom{x}\hspace{48.6ex}@)  areturn
\end{lstlisting}}
\noindent
where the {\tt exp1} and {\tt exp2} are the expression that have compatible type. 
The bytecode example assumes that both {\tt exp1} and {\tt exp2} are variables. The {\tt x} and {\tt y} represent the indexes of {\tt exp1} and {\tt exp2} in the LVT, respectively.

{\bf T13. Mutate Switch Statements.} Mutating the statement following the \texttt{default} label and the expression within the \texttt{case} of switch statement, as well as incorporating the insertion of a mutated \texttt{case} statement derived from neighboring statements.
\fixpattern{
\begin{lstlisting}[language=java,linewidth={\linewidth},basicstyle=\scriptsize\ttfamily]
(@\phantom{x}{\bf \textcolor{darkred}{T13.1}:}\hspace{2ex} \textcolor{red}{-}@)  ...default: stmt1;... (@\phantom{x}\hspace{4.6ex}{\bf T13.1 Bytecode Example:}@) 
(@\phantom{x}\hspace{8.2ex} \textcolor{codegreen}{+}@)  ...default: stmt2;... (@\phantom{x}\hspace{3.7ex} \textcolor{red}{-}@) default: idx_stmt1
(@\phantom{x}{\bf \textcolor{darkred}{T13.2}:}\hspace{2ex} \textcolor{red}{-}@)  ...case exp1: ... (@\phantom{x}\hspace{9.2ex} \textcolor{codegreen}{+}@) default: idx_stmt2
(@\phantom{x}\hspace{8.2ex} \textcolor{codegreen}{+}@)  ...case exp2: ...
(@\phantom{x}{\bf \textcolor{darkred}{T13.3}:}\hspace{2ex} \textcolor{codegreen}{+}@)  ...case exp1: ...
\end{lstlisting}}
\noindent
where {\tt stmt1} is the default statement executed when no cases are matched. {\tt stmt2} is the statement chosen from those following the switch cases. The bytecode examples demonstrate a mutation in the switch statement that replaces the default statement from {\tt stmt1} to {\tt stmt2}. Here, {\tt idx\_stmt1} and {\tt idx\_stmt2} are the indexes of {\tt stmt1} and {\tt stmt2} in constant pool respectively.

{\bf T14. Mutate Fields.} Mutating the field access in an object with other elements (\ie field, method invocation and other expression) that have a compatible type.
\fixpattern{
\begin{lstlisting}[language=java,linewidth={\linewidth},basicstyle=\scriptsize\ttfamily]
(@\phantom{x}{\bf T14.1:}\hspace{2ex} \textcolor{red}{-}@)  ...obj.field1;... (@\phantom{x}\hspace{8.6ex}{\bf T14.1 Bytecode Example:}@)
(@\phantom{x}\hspace{8.2ex} \textcolor{codegreen}{+}@)  ...obj.field2;... (@\phantom{x}\hspace{7.6ex} \textcolor{red}{-}@) getfield #idx_field1
(@\phantom{x}{\bf T14.2:}\hspace{2ex} \textcolor{red}{-}@)  ...obj.field1 = exp;... (@\hspace{0.5ex} \textcolor{codegreen}{+}@) getfield #idx_field2
(@\phantom{x}\hspace{8.2ex} \textcolor{codegreen}{+}@)  ...obj.method(exp);...
(@\phantom{x}{\bf T14.3:}\hspace{2ex} \textcolor{red}{-}@)  ...obj.field1;...
(@\phantom{x}\hspace{8.2ex} \textcolor{codegreen}{+}@)  ...exp...
\end{lstlisting}}
\noindent
where {\tt field1} and {\tt field2} refer to the fields in the buggy code. The {\tt exp} represents an expression that is compatible with the corresponding field's type.
The bytecode example illustrates a case of the T14.1 at the bytecode level. The {\tt \#idx\_field1} and {\tt \#idx\_field2} represent the indexes of {\tt field1} and {\tt field2} in the constant pool, respectively. Note that we utilize {\tt getfield} to retrieve a field from an instance. If we need to retrieve the value of a static field from a class, the opcode should be {\tt getstatic}.

\subsection{Fault Localization}
\label{subsec:fl}
Fault Localization (FL) is a crucial phase for APR and AVR approaches, as it returns a list of suspicious code locations for repair. Existing studies~\cite{liu2019tbar,yuan2020arja,wen2018context} typically utilize off-the-shelf FL tools or directly use the true location of the buggy code (\ie perfect FL). In this study, we use both perfect and spectrum-based FL for \toolname and baselines, as per previous studies~\cite{liu2019tbar,xia2022less,ye2022neural,zhu2021syntax}.
Specifically, the goal of the perfect FL setup is to avoid any potential bias introduced by the FL phase~\cite{liu2019you}, where we provide tools directly with the locations of the vulnerable code.
As for spectrum-based FL setting, we aim to investigate the performance of approaches when using the off-the-shelf FL technique.
For baselines, we use the GZoltar~\cite{campos2012gzoltar} framework with the Ochiai~\cite{abreu2007accuracy} ranking metric to automatically execute test cases and compute the suspicion scores of code lines likely to be the faulty locations.
For \toolname, we first utilize Java Agent~\cite{oracleJavaAgent} to gather instruction coverage information. Subsequently, this data is mapped to the source code line coverage using the {\tt LineNumberTable} extracted from the bytecode files. This table describes the correspondence between the line numbers in the Java source and the offset in the bytecode file. We define a line of source code as covered if any instruction associated with it is executed. We then use the Ochiai ranking metric to measure the suspicious score of code lines, and the corresponding instructions of suspicious code lines are identified for downstream steps.
Note that the FL of \toolname is automated, requiring no human effort. Additionally, while line number information is not essential in Java binaries, our investigation reveals that over 90\% of Java binaries retain the line number information, as discussed in Section~\ref{sec:line_info}.

\subsection{Template Selection}
Based on the templates defined in Section~\ref{subsec:templates}, \toolname determines which templates should be applied to the buggy statements. Unlike template-based APR approaches at the source level~\cite{wen2018context,liu2019tbar, zhang2023gamma}, which typically rely on AST-based matching, it is challenging to construct an AST from bytecode without decompilation~\cite{harrand2020java}. Therefore, \toolname performances a bytecode-based matching approach. 
First, we use the ASM bytecode manipulation framework~\cite{bruneton2002asm} to iterate through all instructions, recording the line numbers associated with each one, as previous studies~\cite{kuleshov2007using,pan2023ppt4j}. We then check all instructions against the set of buggy line numbers provided by the FL phase. If these instructions match a specific template (identified by a particular instruction or a sequence of instructions), we label such matching templates as candidate templates. 
For example, if the opcode of an instruction within buggy instructions (\ie the instructions in buggy lines) is {\tt checkcast}, it matches the {\bf T1. Insert Cast Checker} template. If the opcode of an instruction within buggy instructions is {\tt tableswitch} or {\tt lookupswitch}, it corresponds to the {\bf T13. Mutate Switch Statements} template.
Finally, after inspecting all the suspicious instructions provided by FL phase, we record all candidate templates along with corresponding suspicious locations for the subsequent patch generation phase.

\subsection{Patch Generation and Validation}
In the template selection phase, patches are created by mutating the binaries of the candidate templates. These patched binaries are then run against the test suite. Next, we rank the plausible patches, \ie patches that pass all tests. 
Patches that modify less suspicious locations and change fewer instructions are given higher priority in the ranking.
The reasoning behind this is that correct patches often closely resemble the original program and do not require substantial modifications, as suggested by previous studies~\cite{chen2017contract,wen2018context,le2017s3,wang2020automated}. Finally, the first two authors manually examine all plausible patches to ensure the vulnerabilities are fixed, as per the methodology of previous studies~\cite{yuan2022circle,jiang2021cure, zhang2023gamma, xia2022less}.

Since some programs with vulnerabilities have multiple vulnerable locations, a patch candidate that allows a buggy program to pass a subset of previously failing test cases without failing any previously successful test cases is considered a plausible sub-patch for that program. 
\toolname will continue to generate patches based on sub-patches and assess other patch candidates until all patches are validated or until the time limit is reached.

\subsection{Implementation Details}
We implement \toolname as a practical vulnerability repair tool for Java binaries that without source code.
\toolname is built upon the state-of-the-art bytecode level mutation engine PIT~\cite{coles2016pit}. We have integrated manually transformed bytecode level templates. 
All the experiments were conducted on a server with AMD Ryzen Threadripper 3970X@3.7GHz and 256GB of RAM, running Ubuntu 18.04 LTS, Oracle Java 64-bit Server version 1.7.0\_80 and 1.8.0\_281 (the benchmark requires two versions of Java). 
Regarding the time limitation, we set a four-hour timeout for each vulnerability. This decision is based on the investigation of the test execution time on the Vul4J dataset from Bui et al.~\cite{bui2024apr4vul}.

\section{Study Design}
\label{sec:setup}


\subsection{Research Questions}
In this paper, we seek to answer the following research questions:

{\bf RQ1: Effectiveness of \toolname.} How many real vulnerabilities can be correctly fixed by \toolname?


{\bf RQ2: Efficiency of \toolname.} How doses \toolname perform in terms of efficiency?

{\bf RQ3: Generalizability of \toolname.} What is the generalizability of \toolname in repairing additional real-word vulnerabilities?

\subsection{Settings}
\subsubsection{Baselines.} To answer the research questions, we select 11 APR approaches: ARJA~\cite{yuan2020arja}, Cardumen~\cite{martinez2016astor}, GenProg-A~\cite{yuan2020arja}, jGenProg~\cite{martinez2016astor}, jKali~\cite{martinez2016astor}, jMutRepair~\cite{martinez2016astor}, Kali-A~\cite{yuan2020arja}, RSRepair-A~\cite{yuan2020arja}, TBar~\cite{liu2019tbar}, PraPR~\cite{ghanbari2019practical}, and GAMMA~\cite{zhang2023gamma}. Among them, TBar and PraPR are two state-of-the-art template-based APR tools, and GAMMA is a representative approach for state-of-the-art learning-based APR technique. Additionally, we choose three state-of-the-art learning-based AVR tools, SeqTrans~\cite{chi2022seqtrans}, VulRepair~\cite{fu2022vulrepair} and VulRepair+~\cite{huang2023empirical}.
SeqTrans uses def-use chains to build code sequences that capture syntax and structural details around vulnerabilities with less noise. On the other hand, VulRepair is an automated software vulnerability repair approach based on T5 model~\cite{raffel2020exploring}. It uses a pre-trained model to handle insufficient training data, and it employs the byte-pairs encoding algorithm to reduce the out-of-vocabulary issue.
Huang \etal~\cite{huang2023empirical} conducted an empirical study, proposing modifications to the output format of VulRepair and introducing the ensemble strategy by combining multiple checkpoints. We have incorporated these enhancements into our extended version and named it VulRepairE.
Note that VulRepair and VulRepairE are targeted on C/C++ vulnerabilities and cannot be utilized to repair java vulnerabilities directly. 
To assess their effectiveness, we train both models using the identical dataset employed in SeqTrans by Ponta \etal~\cite{ponta2019manually}. For clarity in representation, we named the versions tailored for Java vulnerabilities as VulRepair-J and VulRepairE-J, denoting the standard and enhanced variants of VulRepair fine-tuned for Java, respectively.

\subsubsection{Terminology.} In APR literature, patches generated by APR tools are classified into three categories: incorrect, plausible, and correct. A patch is considered plausible if it passes all test cases in the given benchmark. If a plausible patch is verified by a human as correctly fixing the bug, then it is classified as a correct patch. Patches that fail to pass either human verification or the test cases are deemed incorrect. In AVR literature, our focus extends beyond functional correctness to include security. Thus, in this study, we categorized the patches into four types: incorrect, plausible, security-fixing, and correct patches. 

\textbf{Incorrect patch.} A patch is considered incorrect if it fails any test case, Proof of Vulnerability (PoV), or human verification.

\textbf{Plausible patch.} In reference to the definition of a plausible patch in the APR domain, we define a plausible patch in the AVR domain as one that can pass all test cases and all PoV checks.

\textbf{Security-fixing patch.} A security-fixing patch is a manually confirmed plausible patch that prevents the exploitation of a vulnerability.

\textbf{Correct patch.} A patch is deemed correct if it passes all test cases, PoV checks, and is manually verified by human.

\subsection{Datasets}
\label{subsec:dataset}
\subsubsection{Dataset for RQ1 and RQ2}
To assess the effectiveness of \toolname fairly, the appropriate dataset is the key requirement. Current dataset mostly are collected from existing database (\eg CVE~\cite{cve2023} and NVD~\cite{nvd2024}). Examples of automatically-created dataset include Big-Vul~\cite{fan2020ac} and CVEfixes~\cite{bhandari2021cvefixes}, while the projectKB~\cite{ponta2019manually} and Vul4J~\cite{bui2022vul4j} are curated manually from the real-world open source software projects. In this study, we select the dataset by the following criteria:

{\bf R1: Real-world Java vulnerability}. The vulnerability project should be open-sourced in Java, and the vulnerability should have an identifier in the CVE database~\cite{cve2023}.

 {\bf R2: Proof of Vulnerability (PoV) exist}. Test-based APR techniques require the test cases (including the failing test cases) to perform the localization of vulnerability, and validate the generated patches. The failing tests (also known as PoV in vulnerability) should fail before repair and pass if the vulnerability is correctly patched.
 
 {\bf R3: Developer patches exist}. Test cases aim to cover as many of the program specifications as possible. However, they are inherently limited and cannot guarantee the correctness of program. Hence, we manually validate the correctness of patches that pass all test cases using developer patches as the ground truth.

 {\bf R4: Vulnerability reproducible}. The source code of the project should be available and compilable. In the vulnerable version, the PoV test cases should fail on the vulnerable version, but pass on patched version.

Based on the above rules, the Vul4J dataset appears to be the only benchmark that satisfies the above requirements. Specifically, although the Big-Vul, CVEfixes and projectKB contain thousands vulnerabilities, most of them have neither PoVs nor corresponding execution environment. Therefore, the Vul4J is used for the evaluation of \toolname and baselines in RQ1 and RQ2. The Vul4J contains 79 vulnerabilities spanning 25 different CWE types from 51 open-source Java projects covering a number of domains including web frameworks and compressed libraries.

\begin{table}[htbp]
  \caption{The additional projects included in the \datasetname dataset}
  \resizebox{0.97\linewidth}{!}{
    \begin{tabular}{lcccc}
    \toprule
        \multicolumn{1}{l}{Project } & \multicolumn{1}{l}{Vuls} & \multicolumn{1}{l}{kLOC} & \multicolumn{1}{l}{Tests} & \multicolumn{1}{l}{CWEs} \\
    \midrule
    jenkinsci/jenkins &  12  &  324 &    3838  &  6\\
    apache/struts &  2 &  359 &  1697  &  2\\
    apache/cxf-fediz &  2  &  60 &  112  & 2 \\
    apache/cxf &  1  & 985  & 6109  &  1\\
    apache/camel &  1  & 958  & 5903   & 1 \\
    apache/activemq & 1   & 447  &   610 &  1\\
    apache/httpcomponents-client & 1   &  104 &  1504  &  1\\
    jenkinsci/ssh-agent-plugin &  1  & 3  &  11  & 1 \\
    nahsra/antisamy & 1   &  7 & 51   &  1\\
    square/retrofit &  1  &  27 &  603   &  1\\
    elastic/elasticsearch &  1  & 696  & 248  &  1\\
    \midrule 
    Average &  2.18  & 361  &  1881  &  1.64\\
    \bottomrule
    \end{tabular}
    }
  \label{tab:vul4jEInfo}%
\end{table}%

\begin{table}[htbp]
  \caption{Number of bugs fixed by \toolname in various projects}
  \resizebox{0.97\linewidth}{!}{
    \begin{tabular}{lccc}
    \toprule
        \multicolumn{1}{l}{Project} & \multicolumn{1}{l}{Plausible} & \multicolumn{1}{l}{Security-Fixing} & \multicolumn{1}{l}{Correct}\\
    \midrule
    alibaba/fastjson & 1    &  1   &   1 \\
    apache/batik  & 1    &  1   &   0 \\
    apache/commons-imaging &    1  &  1  &  1   \\
    apache/struts   &  3   &  2  &  1 \\
    apache/camel & 1    &  1   &   1 \\
    apache/commons-compress &  1   &  1   & 1 \\
    apache/sling-org-apache-sling-xss &  1   &  1   & 1 \\
    apache/tika & 1    &  1   &   0 \\
    cloudfoundry/uaa &  1 &  1   &  1 \\
    jenkinsci/jenkins &   1  &  1   &  0  \\
    javamelody/javamelody &  1   &  1   & 1 \\
    resteasy/Resteasy &  1  &  1  &  1 \\
    spring-projects/spring-framework &  1   &  1    &  1  \\
    spring-projects/spring-security &  1  &  1  &  1 \\
    \midrule
    Total & 16 & 15 & 11 \\
    \bottomrule
    \end{tabular}%
    }
    
  \label{tab:result_projects}%
\end{table}%

\subsubsection{Dataset for RQ3}
In RQ3, we evaluate the generalizability of \toolname by constructing an extended dataset named \datasetname. This dataset adheres to the same selection criteria and data source (\ie ProjectKB) as Vul4J. However, it utilizes a more recent version of ProjectKB, updated until October 2022. It now contains an additional 386 vulnerabilities, increasing the total number from 912 to 1,298. 
Note that the vulnerabilities in ProjectKB are not manually reproduced and verified. Therefore, we construct the \datasetname by following these steps to collect vulnerabilities:

{\bf Step 1}: We first attempt to download the source code from all repositories in ProjectKB, excluding those that are no longer accessible. The vulnerabilities in ProjectKB are manually collected from real world projects that satisfy the R1 criteria.

{\bf Step 2}: We then retain only those vulnerabilities with single-fix commits that modify the Java source code and test code at the same time, since the vulnerability with multiple fix commits may introduce unrelated changes~\cite{bui2022vul4j}. If test code is changed in the same commit, this indicates that the developers added corresponding test cases as a PoV for that vulnerability. We then manually review the changed source code and test code to confirm that these changes actually fix the vulnerability and satisfy criteria R2 and R3.

{\bf Step 3}: Lastly, we filter out vulnerabilities that not compatible with build tools such as Maven, Gradle, and Ant from the vulnerabilities filtered by the previous steps. This is because most of them are challenging to build and compile~\cite{bui2022vul4j}. We then compile and test the vulnerable project both with and without the patches. We only keep the vulnerability if the patched version passes all the test cases, including the PoVs, and the vulnerable version fails the PoVs. In this scenario, we consider the vulnerability to be reproducible and to meet the criteria of R4.

Finally, we construct a dataset, refer to \datasetname, containing 103 vulnerabilities from 55 open source projects. This dataset is approximately 30\% larger than the largest existing dataset, increasing from 79 to 103. As the data source for \datasetname and Vul4J are the same, we present the additional projects included in \datasetname dataset in Table~\ref{tab:vul4jEInfo}. In total, 23 new vulnerabilities were added from 11 different projects, under 11 CWE types. On average, each project has 2.18 vulnerabilities and 1.64 CWE types. Note that 12 of these vulnerabilities come from the {\tt jenkinsci/jenkins} project. This is because this project maintains a rigorous security policy and regularly reports vulnerabilities, often providing PoVs when possible.

\section{Study Result}
\label{sec:eval}
\subsection{Effectiveness of \toolname}
\subsubsection{Experimental Design}
Our first research question focuses on assessing the vulnerability repair performance of fix templates for real-world vulnerabilities. Research by Liu et al.~\cite{liu2019you} suggests that FL techniques can significantly impact the repair performance of various tools. In this context, we aim to evaluate the effectiveness of fix templates. With this mind, we must eliminate any bias introduced by FL. We assume that the vulnerable positions at the statement level are already known. We provide this information directly to the patch generation step of \toolname, rather than obtaining the fault locations from the localization tool.

In addition, we evaluate the effectiveness of \toolname in comparison to existing APR and AVR approaches under perfect FL and spectrum-based FL conditions. The goal of this evaluation under spectrum-based FL is to investigate the effectiveness of \toolname and baselines when using an off-the-shelf FL tool (detailed in~\ref{subsec:fl}).
For the reproduction of APR baselines, we utilize the Cerberus~\cite{shariffdeen2023cerberus} program repair framework to run the APR tools under two FL settings, except PraPR and GAMMA, which are not integrated in Cerberus. For PraPR, GAMMA and AVR baselines, we provide the bug positions at statement level, depending on the input format required by each tool under perfect FL. In the spectrum-based FL setting, we use the FL result from the Cerberus framework to maintain consistency with other baselines.
\begin{table*}[!h]
	\centering
	\scriptsize
	\caption{Vul4J vulnerabilities fixed by fix templates.}
	\label{tab:results_patterns}
	\resizebox{1.0\linewidth}{!} {
    \scriptsize
	\begin{threeparttable}
		\begin{tabular}{l|c|c|c|c|c|c|c|c|c|c|c|c|c|c|c|c|c|c|c|c|c|c|c|c|c|c|c|c|c|c|c|c|c||c|c}
			\toprule
       \multirow{2}{*}{\makecell[l]{{\bf Vul}\\{\bf ID}}} & \multicolumn{2}{c|}{\bf T1} & \multicolumn{5}{c|}{\bf T2} & \multicolumn{3}{c|}{\bf T3} & \multirow{2}{*}{\rotatebox[origin=l]{90}{\bf T4}} & \multicolumn{4}{c|}{\bf T5} & \multicolumn{3}{c|}{\bf T6} & \multicolumn{2}{c|}{\bf T7} & \multicolumn{2}{c|}{\bf T8} & \multicolumn{2}{c|}{\bf T9} & \multicolumn{3}{c|}{\bf T10}& \multirow{2}{*}{\rotatebox[origin=l]{90}{\bf T11}} & \multirow{2}{*}{\rotatebox[origin=l]{90}{\bf T12}} & \multicolumn{3}{c|}{\bf T13} & \multirow{2}{*}{\rotatebox[origin=l]{90}{\bf T14}} & \\\cline{2-11}\cline{13-28}\cline{31-33}
			 & 1 & 2& 1 & 2 & 3 & 4 & 5 & 1 & 2 & 3 & & 1 & 2 & 3 & 4 & 1 & 2 & 3 & 1 & 2 & 1 & 2 & 1 & 2 & 1 & 2 & 3 &  &  & 1 & 2 & 3 & & & \\			\hline
\rowcolor{lightgray}    1& & & & & & & & & & & & & & & & & & & & &\ding{108} & & &\ding{108} & & & & & & & & & & 2/2 & 2/2 \\
                        2& & &\LEFTcircle &\LEFTcircle & & & & & & & & & & & & & & & & & & &\LEFTcircle & & & & & & & & & & & 0/3 & 3/3\\
\rowcolor{lightgray}    4& & & & & & & & & & & & & & & & & & & & &\ding{108} & & & & & & &\ding{108} & & & & & & 2/2 & 2/2 \\
                        6& & & & & & & &\ding{108} & & & &\ding{108} & & & &\ding{108} & & & & & & & & & & & &\LEFTcircle & & & & & & 3/4 & 4/4 \\
\rowcolor{lightgray}    12& & & & & & & &\ding{108} & & & &\LEFTcircle & & &\ding{108} & & & & & &\LEFTcircle & & & & & & &\LEFTcircle & & & & & & 2/5 & 5/5 \\
                        25& & & & & & & & & & & & & & & & & & & & & & &\ding{108} & & & & & & & & & & & 1/1 & 1/1 \\
\rowcolor{lightgray}    33& & & & & & & & & & & & & & & & & & & & & & &\LEFTcircle & & & & & & & & & & & 0/1 & 1/1 \\
                        35& \ding{108} & & & & & & & & & & & & & & & & & & & & & & & & & & & & & & & & & 1/1 & 1/1 \\
\rowcolor{lightgray}    36& & & & & & & & & & & &\ding{109} & & & & & & & & & & &\ding{109} & & &\LEFTcircle & &\LEFTcircle & & & & & & 0/4 & 2/4 \\
                        39& & & & & & & & & & & &\ding{108} & & & & & & & & & & & & &\LEFTcircle & & &\ding{108} & & & & & & 2/3& 3/3 \\
\rowcolor{lightgray}    50& & &\LEFTcircle &\LEFTcircle & & & &\ding{108} & & & & & & & & & & & & &\ding{109} &\ding{109} & & & & & & & & & & & & 1/5 & 3/5 \\
                        55& & & & & & & &\LEFTcircle &\ding{109} & & & & & & & & & & & & & & & & & & & & & & & & & 0/2 & 1/2 \\
\rowcolor{lightgray}    66& & & &\ding{109} & &\LEFTcircle & & & & & &\ding{109} &\ding{109} & & & & & &\ding{109} & & & & & & & & &\LEFTcircle &\ding{109} &\LEFTcircle &\LEFTcircle &\ding{108} & & 1/10 & 5/10 \\
                        71& \ding{108} & & & & & & & & & & & & & & & & & & & & & & & & & & & & & & & & & 1/1 & 1/1 \\
\rowcolor{lightgray}    73& & &\ding{108} &\ding{108} & & & & & & & & & & & & & & & & & & & & & & & & & & & & & & 2/2 & 2/2 \\


\bottomrule
		\end{tabular}
        {\footnotesize \ding{108} indicates that the bug is correctly fixed and \LEFTcircle  means that the bug is security-fixing by the pattern. \ding{109} indicates that the generated patch is plausible but not correct.
        We also present fractions in the penultimate and last column, formatted as x/y. In the penultimate column, x represents the count of fix patterns capable of generating correct patches for a vulnerability, while y denotes the count of fix patterns capable of generating plausible patches for a vulnerability. In the last column, x represents the number of fix patterns that can generate security-fixing patches (including correct ones) for a vulnerability, whereas y denotes the count of fix patterns capable of generating plausible patches for a vulnerability.
        }
	\end{threeparttable}
}
\end{table*}

\subsubsection{Result} Table~\ref{tab:result_projects} represents the effectiveness of \toolname in different projects. Generally, out of 79 vulnerabilities in the Vul4J dataset, \toolname can generate plausible patches for 16 vulnerabilities. It prevents 15 vulnerabilities from being triggered and correctly fixes 11 of them. In previous studies~\cite{bui2024apr4vul}, the APR baseline tool, TBar ~\cite{liu2019tbar}, could securely fix 7 and correctly fix 5 vulnerabilities in Vul4J when assuming perfect localization. While the results of \toolname are promising, we will further examine the effectiveness of each fix template.

Table~\ref{tab:results_patterns} details the correspondence between fix templates and vulnerabilities. Each row indicates the templates that can fix a particular vulnerability, and each column shows the vulnerabilities that a specific fix template can patch. We note that 11 fix templates cannot be used to generate a plausible patch for any vulnerabilities. Five templates (\ie T3.2, T5.2, T7.1, T8.2 and T12) can only generate plausible patches for some vulnerabilities, but none of them are correct or security-fixing patches. 
Note that these fix templates are ineffective in AVR does not necessarily mean that they are worthless, we discuss this in Section~\ref{subsec:effective_pattern}.
Five templates (T2.4, T10.1, T10.2, T13.1 and T13.2) produce security-fixing patches, but none of them are correct. Conversely, 12 templates can generate both security-fixing and correct patches for vulnerabilities. From a single template perspective, T11, which involves removing statements, generates the most useful patches. This template securely fixes four vulnerabilities and correctly resolves two. Template T3.1, involving statement insertion, is most effective at accurately fixing vulnerabilities and correctly repairing three.
From the perspective of a single vulnerability, a vulnerability can be fixed by multiple repair templates. For example, ten templates can create plausible patches for the Vul4J-66 vulnerability. This includes five templates for generating security-fixing patches and one for generating correct patches. The Vul4J-6 vulnerability can also be fixed securely and correctly using four and three fix templates, respectively. This observation highlights the generalizability of repair templates, as templates from different categories can address the same vulnerability.

\subsubsection{Vulnerability Type Analysis}
\begin{table}[htbp]
  \centering
  \caption{\toolname's effectiveness on CWEs.}
  \resizebox{0.98\linewidth}{!}{
    \begin{tabular}{ccccc}
    \toprule
    CWE Type & Name  & \makecell[c]{Security-fixing\\Proportion}  & \makecell[c]{Correct\\Proportion}
 \\
    \midrule
     CWE-20 & Improper Input Validation & 2/9 & 2/9 \\
     CWE-835 & Infinite Loop & 4/8 & 2/8 \\
     CWE-611 & \makecell[c]{Improper Restriction of XML\\External Entity Reference} & 1/6 & 0/6 \\
     CWE-79 & Cross-site Scripting & 2/6 & 2/6 \\
     CWE-77 & Command Injection & 1/1 & 0/1 \\
     CWE-352 & Cross-Site Request Forgery & 1/1 & 1/1 \\
     CWE-200 & \makecell[c]{Exposure of Sensitive Information\\to an Unauthorized Actor} & 1/1 & 1/1 \\
     CWE-522 & Insufficiently Protected Credentials & 1/1 & 1/1 \\
     None$^{*}$  & Not Mapping & 2/13 & 2/13\\
    \bottomrule
    \end{tabular}%
    }
  \label{tab:vul_perf_type}%
  {\footnotesize
    \begin{flushleft} $^{*}$ Approximately 15\% of vulnerabilities cannot be categorized using the current classification scheme~\cite{cwelimit2024}.
    \end{flushleft}
    }
\end{table}%

CWE is a comprehensive catalog that enumerates vulnerabilities weaknesses in software, highlighting potential security issues of varying severity. This enumeration provides valuable guidance to organizations and security analysts, helping them to harden their software systems against potential threats. 
To evaluate the practical implications of Java binary fix templates in real-world scenarios, we perform an investigation into the CWE types of vulnerabilities that can be securely or correctly fixed. 
Table~\ref{tab:vul_perf_type} provides an overview of the effectiveness of \toolname in preventing the exploitation of vulnerabilities across different CWE types. Generally, \toolname can generate patches that securely fix vulnerabilities across eight CWE types and correct patch vulnerabilities across six CWE types in the entire Vul4J dataset.
Specifically, \toolname fixes 50\% of vulnerabilities of the CWE-835 type (\ie Infinite Loop), and correctly repairs 25\% of them. This is because the root cause of CWE-835 is clear, and the key to fixing these vulnerabilities is to break the loop in the right place, making it easier for the fix template to generate the corresponding patches. Apart from frequent CWE types (\eg CWE-20 and CWE-835), \toolname can also generate valid patches for the vulnerabilities of less common CWE types. For example, CWE-200, CWE-352 and CWE-522 each have only one vulnerability in Vul4J, but \toolname manages to fix them all correctly. Furthermore, due to the limitations of the current classification scheme, approximately 15\% of vulnerabilities cannot be categorized~\cite{cwelimit2024}. \toolname also correctly fixes two unclassified vulnerabilities, demonstrating its general applicability.

\subsubsection{Compare with baselines}

\begin{table}[htbp]
  \caption{Comparison with APR and AVR techniques on Vul4J dataset under Perfect FL and Spectrum-based FL}
  \resizebox{0.97\linewidth}{!}{
    \begin{tabular}{ccccccccc}
    \toprule
       \multirow{2}{*}{Tool}  & \multicolumn{2}{c}{Plausible } & \multicolumn{2}{c}{Security-Fixing} & \multicolumn{2}{c}{Correct} & \multicolumn{2}{c}{\%Correct$^{\dagger}$} \\
    \cmidrule(lr){2-3}\cmidrule(lr){4-5}\cmidrule(lr){6-7}\cmidrule(lr){8-9}
         & P$^{*}$  & SB$^{*}$  & P   & SB  & P  & SB  & P &  SB\\
    \midrule
    ARJA        & 8  & 5  & 8  &  5 & 5  & 3  & 62.5 & 60.0 \\
    Cardumen    & 8  & 6  & 3  &  3 & 2  & 2  & 25.0 & 33.3\\
    GenProg-A   & 6  & 4  & 6  &  4 & 2  & 2  & 33.3 &50.0\\
    jGenProg    & 7  & 5  & 4  &  4 & 2  & 2  & 28.6 &40.0\\
    jKali       & 8  & 5  & 5  &  4 & 3  & 2  & 37.5 &40.0\\
    jMutRepair  & 5  & 3  & 2  &  1 & 1  & 1  & 20.0 &33.3\\
    Kali-A      & 7  & 5  & 6  &  4 & 3  & 3  & 42.9 & 60.0\\
    RSRepair-A  & 7  & 5  & 7  &  4 & 5  & 3  & 71.4 &60.0\\
    TBar        & 11 & 7  & 7  &  5 & 5  & 3  &  45.5 &42.9\\
    PraPR       & 9  & 6  & 8  &  5 & 6  & 4  &  66.7 &66.7\\
    GAMMA       & 10 & 6  & 7  &  4 & 6  & 4  &  60.0 &66.7\\
    SeqTrans    & 2  & 2  & 2  &  2 & 2  & 2  & \bf{100.0} &\bf{100.0}\\
    VulRepair-J & 12 & 6  & 8  &  5 & 6  & 5  & 50.0 & 83.3\\
    VulRepairE-J& 12 & 7  & 9  &  6 & 7  & 5  &  58.3 &71.4\\
    \midrule
    \toolname &  \bf{16} &  \bf{9} &  \bf{15}  & \bf{9}   & \bf{11}  & \bf{7} & 68.8 & 77.8 \\
    \bottomrule
    \end{tabular}%
    }
    {\footnotesize
    \begin{flushleft} $^{\dagger}$ \%Correct is the ratio of correct patches over plausible patches. \\
    $^{*}$ P and SB represent the result under perfect FL and spectrum-based FL, respectively.
    \end{flushleft}
    }
  \label{tab:vul4j_PF_FL}%
\end{table}%

Table~\ref{tab:vul4j_PF_FL} presents the effectiveness of state-of-the-art APR and AVR techniques under perfect FL and spectrum-based FL in the Vul4J dataset. Overall, we find that \toolname outperforms other techniques, including both APR and AVR techniques. Under perfect FL, \toolname can securely fix seven more vulnerabilities compared to state-of-the-art ARP tool (\ie GAMMA), and six more than the best-performing AVR tools (\ie VulRepairE-J). This means that \toolname can prevent about 83\% and 40\% more exploits, respectively, than existing state-of-the-art APR and AVR techniques. Looking at the number of correctly fixed vulnerabilities, \toolname generates five and four more correct patches than state-of-the-art AVR and APR techniques, respectively. Compared to the same template-based approaches, TBar and PraPR, which only employ general-bug related templates, \toolname can generate nearly twice as many correct patches.
Under spectrum-based FL, \toolname achieves impressive effectiveness against the baselines, addressing two and three more vulnerabilities than VulRepairE-J and GAMMA, respectively. Furthermore, \toolname outperforms all investigated techniques in terms of the number of security-fixing patches, fixing three more vulnerabilities than VulRepairE-J and four more than GAMMA.
Furthermore, we evaluate the ratio of correct patches to plausible patches (\%Correct). SeqTrans achieved 100\% for both FL settings, indicating that all patches generated by SeqTrans were correct. However, SeqTrans only repaired two vulnerabilities, which accounted for 2.5\% (2/79) of the vulnerabilities in Vul4J.
RSRepair-A attained a 71.4\% correct ratio but only repaired 5 vulnerabilities under perfect FL. Meanwhile, \toolname attains a 68.8\% correct ratio, repairing 11 vulnerabilities. These findings suggest that \toolname can not only repair most vulnerabilities but also maintain a high correctness ratio. 
We have observed that the performance of investigated approaches is inadequate. A possible explanation could be that APR techniques are not tailored for vulnerabilities, and the repair actions for general bugs differ from those for vulnerabilities, as found in previous studies~\cite{bui2024apr4vul,pinconschi2021comparative}. Concerning the state-of-the-art AVR techniques, which are learning-based, the quantity and diversity of historical data directly influence their effectiveness.
In conclusion, both results obtained with perfect FL and spectrum-based FL confirm the superiority of our \toolname over both AVR and APR baselines.

\subsubsection{Overlap Analysis}
\begin{figure}[!t]
    \begin{subfigure}{0.225\textwidth}
      \centering   
      \includegraphics[width=1\linewidth]{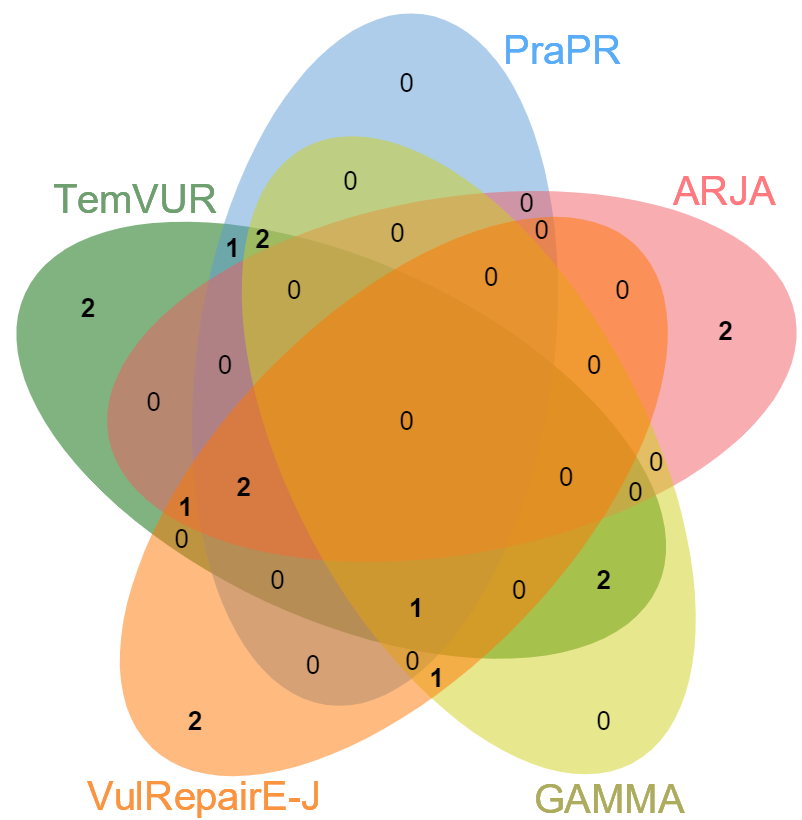}
        \caption{Correct fix}
        \label{fig:overlap_cor}
    \end{subfigure}
    \begin{subfigure}{0.225\textwidth}
      \centering   
      \includegraphics[width=1\linewidth]{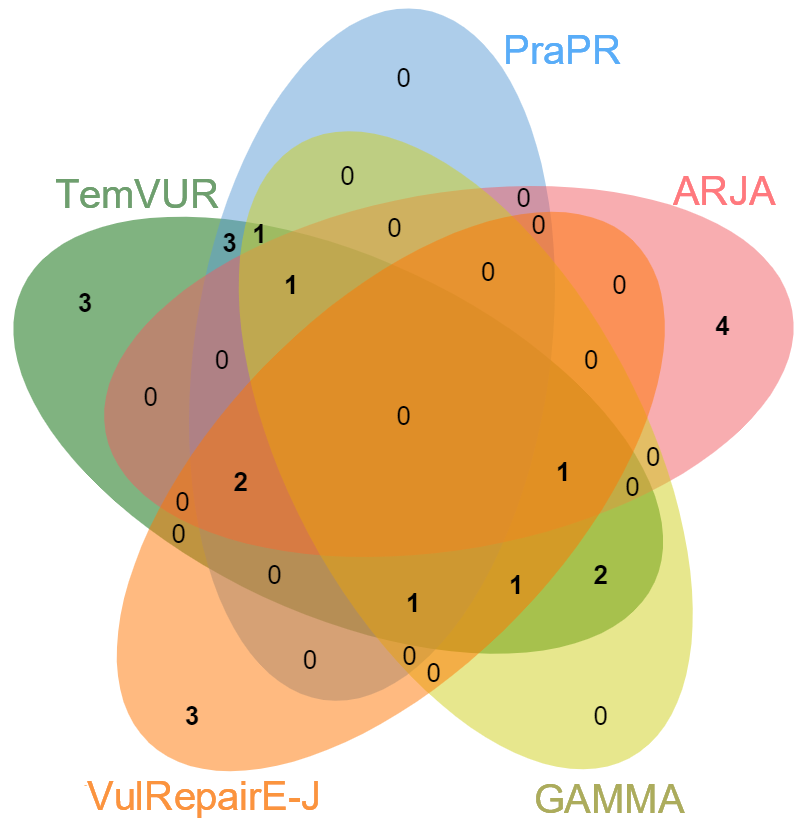}
        \caption{Secure fix}
        \label{fig:overlap_sec}
    \end{subfigure}
  \caption{The overlaps of the vulnerabilities fixed by different approaches.}
  \label{fig:overlap}%
\end{figure}

\begin{table*}[htbp]
  \centering
  \caption{Comparison with baselines on \datasetname}
  \resizebox{1\linewidth}{!}{
    \begin{tabular}{c|ccccccccccccccc}
        \toprule
              & ARJA & Cardumen & GenProg-A & jGenProg & jKali & jMutRepair & Kali-A & RSRepair-A & TBar & PraPR & GAMMA & SeqTrans & VulRepair-J & VulRepairE-J & \toolname \\
        \midrule
        Security-Fixing &   11    &   6    &   6    &   7    &   8    &   4    &   9    &   9    &  10     &   12    &   10    &    2   &  9     &  11  & \bf{20} \\
        Correct         &    2   &    5   &    4   &    4   &    5   &    3   &   5    &    7   &    8   &    9   &    9   &    2  &   7     &  8  & \bf{15} \\
        \bottomrule
    \end{tabular}%
  }
  \label{tab:res_vul4jE}%
\end{table*}%
Different approaches may have unique strengths in addressing different types of vulnerabilities.
This provides the potential to combine existing techniques to achieve improved performance. To evaluate how \toolname complements existing APR techniques, we analyzed overlapping vulnerability fixes using the Vul4J dataset under a perfect FL setting. Specifically, we determined the number of overlapping vulnerabilities addressed by different techniques. 
We selected the best-performing technique from traditional APR, template-based APR, learning-based APR, and AVR. These were ARJA, PraPR, GAMMA, and VulRepairE-J, respectively. As shown in Figure~\ref{fig:overlap_cor}, we found that the \toolname is able to fix two unique vulnerabilities that other techniques fail to repair. Compared with the existing best performing template-based technique PraPR, \toolname can fix more five vulnerabilities that PraPR failed and all PraPR can fix also \toolname can.

We delved into the two vulnerabilities, Vul4J-66 and Vul4J-73, which were uniquely repaired by \toolname. Our examination revealed that the repair actions for these vulnerabilities involved inserting a case statement and a parameter checker, respectively. Despite the clear intentions behind these actions, the investigated approaches failed to fix them. The failure of ARJA, a genetic programming based approach, was due to its inability to generate a population containing the necessary fix ingredients. The learning-based approach, VulRepairE-J, failed because its effectiveness depends on the quantity and diversity of historical data. Our investigation confirmed this, as the training data did not contain similar repair actions.
Regarding PraPR, a template-based approach, its failure to address these vulnerabilities stemmed from a lack of corresponding templates. Since its templates are designed for general bug repair, this highlights the importance of vulnerability-specific templates for handling such scenarios.

The overlap of vulnerabilities securely fixed by selected techniques is depicted in Figure~\ref{fig:overlap_sec}, which reveals a similar phenomenon. Specifically, \toolname uniquely prevents the exploitation of three vulnerabilities, while ARJA and VulRepairE-J prevent four and three vulnerabilities, respectively.

The overlap analysis between \toolname and existing techniques shows that \toolname not only achieves promising results, but also uniquely repairs some vulnerabilities. This suggests that \toolname can serve as a suitable baseline for future research to identify vulnerabilities that can be resolved with a straightforward approach.
Moreover, we observed that different types of techniques have their own strengths, suggesting that an ensemble of existing approaches could be promising~\cite{zhong2024practical}.

\begin{table}[htbp]
  \centering
  \caption{The Efficiency of \toolname and Baselines on Vul4J Dataset}
    \begin{tabular}{lrrrr}
    \toprule
          & Avg(s) & Min(s) & Max(s) \\
    \midrule
    ARJA  & 5,240.5 & 23.7  & 14,400.0 \\
    Cardumen & 1,816.7 & 10.3  & 11,341.6 \\
    GenProg-A & 10,168.1 & 50.4  & 14,400.0 \\
    jGenProg & 3,829.6 & 14.5  & 14,400.0 \\
    jKali & 232.4 & 10.3  & 3,847.5 \\
    jMutRepair & 170.9 & 13.0  & 2,179.1 \\
    Kali-A & 1,148.4 & 40.5  & 14,400.0 \\
    RSRepair-A & 7,422.5 & 17.8  & 14,400.0 \\
    TBar  & 581.8 & 23.2  & 14,400.0 \\
    PraPR & 57.8  & 7.6   & 1,756.5 \\
    GAMMA & 6,779.7 & 29.3  & 14,400.0 \\
    SeqTrans & 207.1 & 31.0  & 4,572.6 \\
    VulRepair & 2,556.6 & 35.2  & 14,400.0 \\
    VulRepair+ & 2,427.9 & 28.3  & 14,400.0 \\
    \midrule
    \toolname & 61.5  & 8.8   & 2,087.3 \\
    \bottomrule
    \end{tabular}%
  \label{tab:time_efficiency}%
\end{table}%
\subsection{Efficiency of \toolname}
\subsubsection{Experimental Design} 
Based on research~\cite{chen2017contract}, compiling and validating of patches can consume up to 92.8\% of the total execution time of automated program repair tools. Repairing vulnerabilities at the binary code level allows the code to be executed directly without compilation. In this research question, we aim to evaluate the efficiency of \toolname and its baselines by measuring the running time on the Vul4J dataset. Note that for learning-based techniques (\ie GAMMA, SeqTrans, VulRepair-J, and VulRepairE-J), we measure only the time cost of patch generation and validation, excluding the time cost of the training phase. All tools are executed on the same machine using a single thread.
To mitigate any bias introduced by FL, we conduct the experiment in this research question only under perfect FL conditions. 

\subsubsection{Result}
Table~\ref{tab:time_efficiency} shows the time cost (in seconds) of \toolname and baselines on all Vul4J bugs under perfect FL. 
In summary, we observed that two binary code level approaches (\ie \toolname and PraPR) were significantly more efficient than other approaches. In comparison to the template-based approach on the source code, TBar, \toolname requires only a tenth of the time required by TBar, while fixing about twice as many vulnerabilities. PraPR, on the other hand, only implements templates for APR rather than AVR, and thus requires less time but fixes fewer vulnerabilities compared to \toolname.
We also found that each type of approach has distinct characteristics in terms of efficiency.
Specifically, tools based on genetic programming (\ie ARJA, GenProg-A, jGenProg, and RSRepair-A) are less efficient, costing thousands of seconds to repair each vulnerability on average. For instance, GenProg-A takes about ten thousand seconds on average to repair a vulnerability. 
This inefficiency could be due to the quadratic time complexity of genetic programming, which make it challenging to apply it to large-scale problems~\cite{lobo2000time}.
We observe that jKali and JMutRepair take on average only about two hundred seconds to repair a vulnerability. We infer that this is due to their simple patch generation rules, which result in a relatively small search space. For instance, the primary patch generation rules implemented in jKali consist of removing statements, modifying if conditions to `true' or `false', and inserting return statements. 
Regarding learning-based techniques (\ie GAMMA, SeqTrans, VulRepair-J and VulRepairE-J), we observed that the time efficiency varied, ranging from approximately seven hundred seconds to about seven thousand seconds. Upon examination of the results, we found that each learning-based approach generally took a similar amount of time to generate patches, but differed in patch validation. Specifically, on average, GAMMA required about 22 seconds to generate patches per vulnerability, wheresa SeqTrans took around 2 seconds per vulnerability. This suggests that a significant amount of time is spent compiling and validating, which is directly correlated to the number of generated patches. The disparity in the number of patches produced by GAMMA and SeqTrans supports this observation: on average, GAMMA generates 475.8 patches per vulnerability, whereas SeqTrans generates only 10 patches per vulnerability.
Furthermore, template-based approaches (\ie TBar, PraPR, and \toolname) maintain high effectiveness and efficiency. For instance, both TBar and ARJA successfully repair five vulnerabilities in Vul4J under perfect FL conditions, yet TBar accomplishes this task in only about a tenth of the time required by ARJA. This underscores the pivotal role of templates in guiding patch generation.
These findings underscore the promise of binary code-level AVR in assisting both vendors and users in mitigating the adverse effects of increasing vulnerabilities.

\subsection{Generalizability of \toolname}
\subsubsection{Experimental Design} We have shown the impressive performance of \toolname on the widely-used Vul4J dataset. The research from Durieux \etal~\cite{durieux2019empirical} indicates a common dataset overfitting phenomenon in program repair tools evaluation. This means that tools may perform better on certain datasets. To avoid potential overfitting in Vul4J and to assess the general applicability of \toolname, we conduct an extended investigation on \datasetname, which is extended from Vul4J but approximately 30\% larger than Vul4J. We expect that using a larger dataset with more vulnerabilities will help mitigate the impact of overfitting.

\subsubsection{Result}
Table~\ref{tab:res_vul4jE} presents the comparison result of \toolname against baselines on the \datasetname dataset. Overall, \toolname manages to correctly fix 20 and 15 vulnerabilities respectively, outperforming both existing APR and AVR approaches. We find that the performance of \toolname on \datasetname is roughly similar to its performance on the Vul4J dataset. Specifically, \toolname securely fixes 19.0\% (15/79) of the vulnerabilities in \datasetname. This is nearly the same performance as the 19.4\% (20/103) achieved in the Vul4J dataset. Similarly, 14.6\% (15/103) of the vulnerabilities in \datasetname are fixed correctly and in the Vul4J dataset it is 13.9\% (11/79). 
VulRepairE-J securely and correctly repairs 10.7\% (11/103) and 7.8\% (8/103) of the vulnerabilities in \datasetname respectively. This performance is lower than in Vul4J, where it securely repairs 11.4\% (9/79) and correctly repairs 8.9\% (7/79) of the vulnerabilities. Such performance degradation also occurs with other learning-based techniques such as VulRepair-J and SeqTrans. The possible reason for this reduction in performance is that \datasetname contains a wider variety of project types, and the existing training data for such learning-based approaches is limited. 
Note that GAMMA maintains significant effectiveness in \datasetname, as it is a cloze-style APR technique that masks the vulnerable part and requires the tool to fill it. It leverages the pre-training model to generate the patch without training on historical vulnerability-fixing data.
Taken together, these phenomena suggest that \toolname is not overfitted to a specific dataset and is thus generalizable. Moreover, since \toolname does not require a training phase, it is less likely to encounter the generalizability problem common in learning-based tools, which heavily rely on the diversity of their training data.

\section{Discussion}
\label{sec:dis}
\subsection{Limitations of \toolname}
\label{subsec:effective_pattern}
\begin{figure}
    \centering
\begin{lstlisting}[language=diff,frame=lines,linewidth={\linewidth},basicstyle=\footnotesize\ttfamily,aboveskip=8pt]
 if (xmlIn == null) {
     xmlIn = XMLInputFactory.newInstance();
+    xmlIn.setProperty(XMLInputFactory.IS_SUPPORTING_EXTERNAL_ENTITIES, Boolean.FALSE);
 }
\end{lstlisting}
    \caption{The patch from the developer for the Vul4J-46 vulnerability.}
    \label{fig:vul_46}
\end{figure}

\begin{figure}
    \centering
\begin{lstlisting}[language=diff,frame=lines,linewidth={\linewidth},basicstyle=\scriptsize\ttfamily]
 public String encodeForJSString(String source) {
-    return source == null ? null : Encode.forJavaScriptSource(source);
+    return source == null ? null : Encode.forJavaScript(source).replace("\\-", "\\u002D"); (@\textcolor{codegray}{// Developer's patch}@)
+    return source == null ? null : Encode.forJavaScript(source);(@\textcolor{codegray}{// \toolname's patch}@)
}
\end{lstlisting}
    \caption{The patches from the developer and \toolname for the Vul4J-23 vulnerability.}
    \label{fig:vul_23}
\end{figure}

According to the findings in RQ1, some templates cannot generate plausible patches for any vulnerabilities, and many vulnerabilities in Vul4J cannot be repaired correctly. After investigating the failures, we identified three possible reasons to explain this.

{\bf Inadequate Dataset}. The Vul4J dataset does not contain a sufficient number and types of vulnerabilities. It contains only 79 vulnerabilities spanning 25 different CWE types. This is insufficient compared to the approximately 228k records in the CVE database~\cite{cve2023} and the 938 CWE types~\cite{cwe2024} as of February 1, 2024.
    
{\bf Inefficient Search}. Searching for donor code can be inefficient when trying to find relevant code to apply these patterns. Figure~\ref{fig:vul_46} shows the developer's patch for the Vul4J-46 vulnerability\footnote{\url{https://github.com/FasterXML/jackson-dataformat-xml/commit/f0f19}}. This patch inserts a statement that sets the property of the object {\tt xmlIn}. Theoretically, this could be repaired by template T3.1 (\ie Insert statements) since a similar statement exists in another class under the same directory. However, due to the large search space and inefficient search strategy, the \toolname failed to find useful donor code to apply this template.
    
    
{\bf Partial Repair}. Some fix templates can partially repair vulnerabilities, preventing them from being triggered in certain situations, but cannot completely resolve the vulnerabilities. Figure~\ref{fig:vul_23} represents both the developer-written patch\footnote{\url{https://github.com/apache/sling-old-svn-mirror/commit/7d236}} and the patch from \toolname. This vulnerability is categorized as CWE-79 (\ie Cross-site Scripting), which allows the attacker's input to be executed as a script. In this case, the method {\tt forJavaScriptSource} encodes the input, but it is not restrictive enough that some input patterns allow script tags to pass through unencoded. The developer-written patch replaces the {\tt forJavaScriptSource} with a more restrictive function, {\tt forJavaScript}, and replaces ``{\tt \textbackslash -}'' with ``{\tt \textbackslash u002D}'' to prevent it from being interpreted as a special character by the JavaScript interpreter. 
The \toolname's patch also replaces {\tt forJavaScriptSource} with {\tt forJavaScript}, which escapes most harmful characters that could potentially cause unauthorized actions, except for the dash (\ie ``{\tt \textbackslash -}'').
A dash can be used in JavaScript for subtraction operations or to denote a comment, both of which could be used in an XSS attack vector.
Although this patch does not completely repair the vulnerability, it prevents most malicious input, significantly reducing the risk to downstream users.

\subsection{Complementarity Between Source-Level and Binary-Level AVR}
\label{sec:limitation}
\begin{figure}[!t]
    \begin{subfigure}{0.45\textwidth}
      \centering   
      \includegraphics[width=1\linewidth]{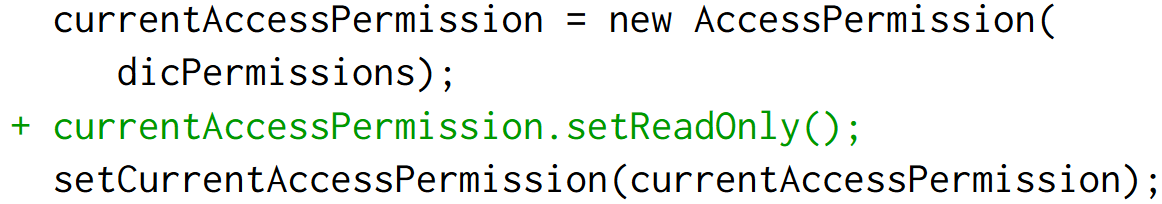}
        \caption{Developer's source code patch for Vul4J-59}
        \label{fig:vul4J_59_dev}
    \end{subfigure}
    \begin{subfigure}{0.2\textwidth}
      \centering   
      \includegraphics[width=1\linewidth]{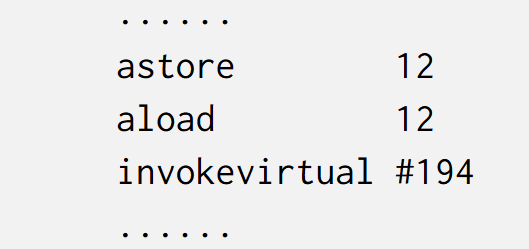}
        \caption{Donor bytecode}
        \label{fig:donor_code_59}
    \end{subfigure}
    \begin{subfigure}{0.2\textwidth}
      \centering   
      \includegraphics[width=1\linewidth]{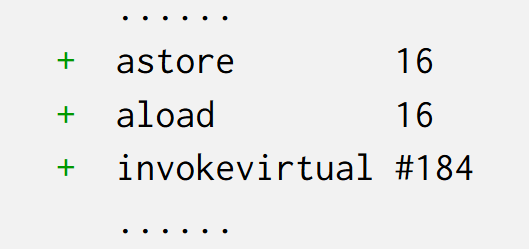}
        \caption{Bytecode patch}
        \label{fig:oracle_patch_vul59}
    \end{subfigure}
  \caption{The patches and donor code for Vul4J-59.}
  \label{fig:vul4j_59}%
\end{figure}

While \toolname delivers promising results for repairing Java vulnerabilities at the binary level compared to tools that repair vulnerabilities at the source level, it still has some limitations. This is primarily because we do not have access to extensive information, such as detailed typing and context information, which makes some simple repair actions difficult at the binary level.
For example, the Vul-59 vulnerability is successfully repaired by several source code level tools, such as ARJA and RSRepair, but not by \toolname. Figure~\ref{fig:vul4J_59_dev} illustrates the developer's patch for Vul4J-59. This vulnerability can be fixed by inserting a statement that invokes the {\tt setReadOnly} function, which can be found in a class in the same folder. This allows source-level tools to easily reuse the donor code. However, for \toolname, which operates at the binary level, finding and reusing the donor code is challenging because the same source code snippet might differ when compiled into binaries. Specifically, Figure~\ref{fig:donor_code_59} and Figure~\ref{fig:oracle_patch_vul59} depict the donor bytecode and bytecode patch for Vul4J-59, respectively. We find that the added bytecode in the bytecode patch is not identical to the donor bytecode due to differences in the local variable table, constant pool, and so on. Therefore, while \toolname outperforms existing source-level AVR approaches, the latter still have their advantages. This situation underscores the potential of ensemble tools that target at different levels to achieve better performance~\cite{zhong2024practical}.

\subsection{Line Number Information in Java Binaries}
\label{sec:line_info}
The fault localization phase of \toolname operates at the source code line level, requiring line number information that maps bytecode instructions to corresponding source code lines. However, this line number information, part of the debug information, may be stripped from binaries. To evaluate how many Java binaries retain line number information, we conducted an empirical study. We collected the 5000 most popular Java binaries from the Maven central repository, based on the number of usage. Our findings showed that over 90\% of class files in these binaries contained line number information, a result that aligns with a recent study~\cite{pan2023ppt4j}. Hence, we believe that retaining line number information is common practice. This makes \toolname applicable in many practical scenarios.

\subsection{Threats to Validity}
\label{subsec:threats}
{\bf Internal Threats.}
The first internal threat arises from the manual validation of patch correctness. To mitigate the potential bias, two co-authors independently verify the correctness of plausible patches, following previous studies~\cite{yuan2022circle,jiang2021cure, zhang2023gamma, xia2022less}. A patch is deemed correct if both authors find it semantically equivalent to a ground truth patch. A security-fixing patch is a manually confirmed plausible patch that prevents the vulnerability from being exploited.
To facilitate replication and verification of our experiments, we have made the relevant materials, including source code and correct patches, publicly available in our online repository.

The second threat arises from our practice of manually inspecting only those patches that pass all test suites (\ie plausible patches), in line with common practice in the community~\cite{bui2024apr4vul,zhang2023gamma,liu2019tbar}. This could lead to an unfair evaluation of some patches. For instance, certain patches might be correct or security-fixing but fail to pass the test cases due to some test cases are too rigid. For example, certain tests\footnote{\url{https://github.com/apache/struts/commit/23743}} expect to receive a certain message and throw a failure if they do not. However, we believe this is fair since all tools under investigation are evaluated using the same test suites.

{\bf External Threats.}
The main external threat to validity comes from the chosen evaluation benchmarks. The performance claims for \toolname may not apply directly to other datasets. Since Vul4J was the only dataset that met our selection criteria, we tried to reduce this threat by creating \datasetname and evaluating the generalizability of \toolname through the evaluation on \datasetname. The results show that \toolname maintains similar effectiveness on both Vul4J and \datasetname, confirming its general applicability across different datasets.

\section{CONCLUSION}
\label{sec:conc}
This paper presents \toolname, a straightforward yet effective template-based approach for repairing vulnerabilities in Java binaries.
Drawing on insights from the existing literature, it substantially mitigates vulnerabilities, achieving a 57.1\% improvement over state-of-the-art approaches as shown by evaluation on the Vul4J dataset, highlighting its effectiveness in reducing risk. Besides, the introduction of the \datasetname dataset broadens the scope of the evaluation and further validates \toolname's effectiveness across various vulnerabilities.
In summary, \toolname offers a promising solution for repairing the Java vulnerability without source code, filling crucial gaps in current practices and bolstering software security. Crucially, \toolname also serves as a baseline to observe the performance attainable through the direct application of repair templates and to illustrate the enhancements provided by future approaches. 
All data in this study are publicly available at:
\textbf{\url{https://zenodo.org/records/11084782}}.

\ifCLASSOPTIONcaptionsoff
  \newpage
\fi

\ifCLASSOPTIONcompsoc
  \section*{Acknowledgments}
\else
  \section*{Acknowledgment}
\fi


\ifCLASSOPTIONcaptionsoff
  \newpage
\fi



%
\bibliographystyle{IEEEtran}
\bibliography{bib/references}
%







\end{document}